\documentclass[prd,twocolumn,showpacs,nofootinbib,amsmath,amssymb,superscriptaddress]{revtex4-1}

\usepackage{graphicx}
\usepackage{dcolumn}
\usepackage{bm}
\usepackage{amsmath,amssymb,bbm,mathrsfs,nicefrac}
 \usepackage{array,multirow}
 \usepackage{float}
\usepackage{tikz}
\usepackage{amsmath}
\usepackage{amssymb}
\usepackage{graphicx,textcomp,float,gensymb,wrapfig, enumitem,comment,dsfont,framed,slashed,appendix,wrapfig,xcolor}
\usepackage{ bbold }
\usepackage{braket} 
\usepackage{ mathrsfs }
\usepackage{etoolbox,hyperref,makecell}
\usepackage{feynmp}
\usetikzlibrary{arrows.meta}
\usepackage{empheq}
\usepackage[normalem]{ulem}
\usepackage{scrextend}
\usepackage{slashed}
\newcommand{\bea}{\begin{eqnarray}}  
\newcommand{\eea}{\end{eqnarray}}

\newcommand{\vev}[1]{\langle #1 \rangle}

\definecolor{darkgreen}{rgb}{0.2, 0.3, 0.1}

\usepackage[utf8]{inputenc}

\begin{document}

\title{
Mirror Neutron Stars
}

\author{Maur\'icio Hippert}
\affiliation{Illinois Center for Advanced Studies of the Universe, Department of Physics, University of Illinois at Urbana-Champaign, Urbana, IL 61801, USA}

\author{Jack Setford}
\affiliation{Department of Physics, University of Toronto, Toronto, ON M5S 1A7, Canada}

\author{Hung Tan}
\affiliation{Illinois Center for Advanced Studies of the Universe, Department of Physics, University of Illinois at Urbana-Champaign, Urbana, IL 61801, USA}

\author{David Curtin}
\affiliation{Department of Physics, University of Toronto, Toronto, ON M5S 1A7, Canada}

\author{Jacquelyn Noronha-Hostler}
\affiliation{Illinois Center for Advanced Studies of the Universe, Department of Physics, University of Illinois at Urbana-Champaign, Urbana, IL 61801, USA}

\author{Nicol\'as Yunes}
\affiliation{Illinois Center for Advanced Studies of the Universe, Department of Physics, University of Illinois at Urbana-Champaign, Urbana, IL 61801, USA}

\date{\today}

\begin{abstract}

The fundamental nature of dark matter is entirely unknown. A compelling candidate is Twin Higgs mirror matter, invisible hidden-sector cousins of the Standard Model particles and forces. This predicts mirror neutron stars made entirely of mirror nuclear matter. We find their structure using realistic equations of state, robustly modified based on first-principle quantum chromodynamic calculations, for the first time. This allows us to predict their gravitational wave signals, demonstrating an impressive discovery potential and ability to probe dark sectors connected to the hierarchy problem.


\end{abstract}

\maketitle
\onecolumngrid

\tableofcontents

\newpage

\twocolumngrid


\section{Introduction}

Dark Matter makes up $\sim 85\%$ of matter in our universe, 
yet its properties remain unknown. We only have gravitational information on large astrophysical or cosmological scales:  how much exists, modest bounds on self-interactions and more stringent bounds on  interactions with Standard Model (SM) matter~\cite{Tulin:2017ara}. 
These constraints still allow for an enormous space of possibilities. 
Uncovering the fundamental nature of dark matter would solve one of the greatest mysteries of modern science.

It is possible all of dark matter is a single, collisionless particle species.
This possibility is appealing not only for its simplicity, but also because such dark matter candidates occur in predictive theories of particle physics that address an \textit{a priori} unrelated issue.
For example, the \emph{Hierarchy Problem} is the inconsistency between the Higgs mass and the value expected from quantum field theory.
In the SM, the electroweak scale and masses of \emph{fundamental} particles are set by the mass of the Higgs boson. 
However, 
the Higgs mass is sensitive to unknown physics at tiny distance scales, including details of quantum gravity.
Quantum corrections should set the Higgs mass $m_H$ near the Planck Scale $\sim 10^{18}$ GeV, which is in direct conflict with the measured value $m_H \approx 125$~GeV.

Supersymmetry (SUSY)~\cite{Martin:1997ns} solves this problem by introducing particles that 
cancel quantum corrections to $m_H$. These might include Weakly Interacting Massive Particles (WIMPs) that could be dark matter. 
This would imply a relationship between dark matter signatures and the observed dark matter relic abundance. 
Unfortunately, direct detection signals of WIMPs and collider signals of SUSY have not yet been found~\cite{ATL-PHYS-PUB-2020-020, Kang:2018odb}, straining this hypothesis.

The WIMP paradigm is theoretically attractive, but paints a picture of dark matter in stark contrast to our experience with SM matter.  SM matter contains various particles that interact via four fundamental forces -- why should dark matter be so simple in comparison? 
Nothing prevents different dark particles, endowed with their own dark forces (see e.g.~\cite{Alexander:2016aln}). However, dark sectors of similar complexity to the SM, known  as \emph{Dark Complexity}~\cite{Curtin:2019ngc, Curtin:2020tkm}, are underexplored.

Given the lack of evidence for theories like WIMPs or SUSY, the plausibility of complex dark matter, and its wildly different experimental signatures compared to minimal dark matter, it is imperative to understand the physical consequences of Dark Complexity. 
This seems daunting given the vast space of possibilities, but 
unsolved fundamental puzzles of high-energy particle physics can again provide guidance.

A well-motivated example of Dark Complexity is \emph{mirror matter}~\cite{Berezhiani:1995am,Chacko:2005pe}, a dark matter candidate related to SM matter via a discrete symmetry. 
Part of dark matter would be made up of mirror protons, neutrons, and electrons, which do not interact under SM gauge forces, making them invisible. Instead, mirror matter interacts with itself via mirror versions of electromagnetism, the weak and strong forces. 
Just as WIMPs occur in supersymmetry, mirror matter is a prediction of another solution to the Hierarchy Problem, the \emph{Twin Higgs mechanism}~\cite{Chacko:2005pe, Chacko:2016hvu, Chacko:2018vss}.
Twin Higgs predicts different collider signatures than  SUSY~\cite{Burdman:2014zta,Craig:2015pha},  evading existing constraints \cite{ATL-PHYS-PUB-2020-020} from LHC searches.
Twin Higgs mirror matter is an excellent starting point to understand Dark Complexity. Several of its predictions have already been theoretically studied and will soon be tested \cite{Chacko:2016hvu, Chacko:2018vss, Curtin:2019ngc, Curtin:2020tkm}.

We examine for the first time one of the simplest, most spectacular predictions of the Twin Higgs: if dark matter contains mirror matter, then the cosmos should be filled with \emph{mirror neutron stars},  invisible degenerate stellar objects made \emph{entirely} of mirror nucleons, analogous to SM neutron stars. 
These objects are a generic prediction of non-minimal dark matter theories with a confining force similar to SM quantum chromodynamics (QCD), but have never been studied before in realistic detail. 
Previous investigations studied other kinds of exotic compact objects~\cite{Cardoso:2019rvt}; used SM neutron stars as probes of dark sectors (e.g.~\cite{Ellis:2018bkr}); or considered neutron-star-like objects that were composed of exact copies of SM nuclear matter~\cite{Beradze:2019yyp}, fundamental dark fermions with simple interactions~\cite{Narain:2006kx, Kouvaris:2015rea, Gresham:2018rqo}, or highly simplified general dark nuclear matter~\cite{Maedan:2019mgz}. 
Ours is the first realistic study of mirror neutron star properties arising from a dark QCD sector that is different from the SM. This allows us to predict their mass-radius relationship, Love Numbers and moments of inertia in detail, explicitly demonstrating for the first time that gravitational wave detectors can discover these objects using standard analysis techniques, probe their dark sectors and uncover their connection to the Hierarchy Problem.

\section{Twin Higgs Mirror Matter}


The Mirror Twin Higgs mechanism \cite{Chacko:2005pe, Chacko:2016hvu}  addresses the Hierarchy Problem by introducing a hidden-sector copy of the SM particles and forces. As in the SM, mirror quarks form mirror protons and neutrons, coalesce into mirror nuclei, and form mirror atoms with mirror electrons~\cite{Chacko:2018vss}.
The symmetry relating the SM and mirror sector means that the baryogenesis mechanism that generates the matter-excess of our universe should have a mirror-sector analogue, meaning a significant fraction of DM could be in the form of mirror matter.
Mirror matter and SM matter must interact very faintly via the Higgs boson, but  for the purposes of our astrophysical discussion we can regard mirror matter as invisible and only detectable gravitationally.  

Earlier studies of mirror matter focused on an \emph{exact} copy of SM matter~(e.g. \cite{Berezhiani:1995am}).
In contrast, the minimal Mirror Twin Higgs predicts mirror matter that is a SM copy, except that the vacuum expectation value  $f$ of the mirror Higgs is scaled up compared to the value for the SM Higgs $v$ by a factor of $f/v \sim 3 - 10$. 
The mirror quarks, leptons, and gauge boson masses are scaled by the same factor.
The lower bound $f/v \gtrsim 3$ ensures agreement with experimental constraints, including LHC Higgs coupling measurements~\cite{Aad:2019mbh}. Large values of $f$ reduce the theory's ability to solve the Hierarchy Problem, making $f \lesssim 10$ a reasonable upper bound. 
Properties of Twin Higgs mirror matter significantly deviate from the SM, while being similar enough to allow for robust predictions.

Just like SM matter, mirror matter would display rich dynamics at all size scales. 
In the early universe, the presence of mirror matter gives rise to cosmological signatures like mirror-baryo-acoustic oscillations and faint dark radiation signals in the Cosmic Microwave Background~\cite{Chacko:2016hvu, Chacko:2018vss}. More general scenarios may produce gravitational waves from the dark hadronization phase transition~\cite{Schwaller:2015tja}.
Each galaxy would likely contain a mirror-baryon-component and form Mirror Stars~\cite{Curtin:2019ngc} that could be detectable, depending on unknown aspects of mirror stellar physics and the feeble coupling between SM and mirror matter.
Details of these dynamics are uncertain, depending on the precise parameters of  mirror matter. However, the endpoint of mirror-stellar-evolution is, as for SM matter, relatively simple:  compact objects like neutron stars, white dwarfs, or black holes. 
Mirror neutron stars are a generic prediction of Dark Complexity and we investigate  their signatures by studying the Twin Higgs benchmark model.

At low energies, mirror QCD behaves similarly to SM QCD, except that the strong confinement scale $\Lambda_\mathrm{QCD}'$ is larger than the SM (where $\Lambda_\mathrm{QCD} \approx 250$ MeV) 
by a factor of $\Lambda_\mathrm{QCD}'/\Lambda_\mathrm{QCD} \approx 1.3\  (1.7)$ for $f/v = 3\ (10)$~\cite{Chacko:2018vss}. 
The mirror neutron and proton masses increase accordingly.
The pion mass is related to the fundamental quark masses and confinement scale, $m_\pi \propto \sqrt{\Lambda_\mathrm{QCD} m_q}$, making the mirror pion 2 (4) times heavier than $m_\pi^{SM}=140$ MeV for $f/v = 3\ (10)$.
This allows us to predict the properties of mirror neutron stars.

Our work focuses on this symmetric Twin Higgs benchmark model, but the results depend mainly on the dark confinement scale. This also provides insight into other mirror-QCD scenarios with different mirror hadron spectra~\cite{Barbieri:2016zxn}.

\subsection{Hierarchy Problem and Neutral Naturalness}

The idea that a subcomponent of dark matter could consist of particles with qualitatively similar properties to those in the Standard Model (SM) has been around for a long time~\cite{Blinnikov:1982eh, Blinnikov:1983gh, Kolb:1985bf, Glashow:1985ud, Khlopov:1989fj, Foot:1995pa,Berezhiani:1995am,  Chacko:2005pe, Chacko:2016hvu, Craig:2016lyx, Chacko:2018vss} (see also e.g.~\cite{Roux:2020wkp}). In particular, mirror models proposed the existence of a mirror-symmetric sector with mirror analogues of all of the SM particles and gauge groups. Since these new particles interact under their own copy of the SM $SU(3)_c \times SU(2)_L \times U(1)_Y$ gauge group, their interactions with ordinary matter take place via portal interactions. Portal interactions couple fields from one sector to fields from another  (such as the vector portal interaction that introduces a small mixing of the SM photon and mirror photon). Such interactions do not violate any symmetries and are expected to be present at some level, but they can be almost arbitrarily small, which explains why this new sector of matter and forces may have remained hidden so far.

More recently, there has been renewed interest in mirror sectors due to their relevance to the Hierarchy Problem~\cite{Chacko:2005pe}. The Hierarchy Problem refers to the extreme hierarchy between the scale of electroweak interactions and the scale of gravity -- specifically the smallness of the Higgs boson mass compared to the Planck scale \cite{Weinberg:1975gm}. The Higgs mass is the only dimensionful parameter in the SM Lagrangian, and is unprotected from large quantum corrections from new physics at high energy scales, which must finely cancel with the bare Higgs mass parameter. This extreme fine tuning can be ameliorated in Beyond Standard Model (BSM) theories which introduce new symmetries that protect the Higgs mass from high-scale quantum corrections, or make the Higgs a composite state that only exists at low energies. The most well-known examples of such models are Supersymmetry \cite{Martin:1997ns} and Composite Higgs \cite{Bellazzini:2014yua}. Both of these classes of models predict the existence of new particles charged under QCD with masses not much higher than a TeV, and they were part of the motivation for the construction of the Large Hadron Collider (LHC). However, in the absence of any discovery of BSM physics at the LHC \cite{ATL-PHYS-PUB-2020-020}, such models are under increasing tension. An alternative class of theories has received increased attention in recent years, known as \emph{neutral naturalness}, in which the lightest BSM particles are all charged under a \emph{hidden} gauge group and whose production rates at the LHC are significantly suppressed as a result~\cite{Chacko:2005pe, Burdman:2006tz, Cai:2008au}. These models are able to address the Hierarchy Problem when this hidden sector is related to the SM via some symmetry, which can result in approximate cancellations in the contributions to the mass of the Higgs. 
Neutral naturalness therefore motivates particular types of mirror matter.  

The simplest realisation of neutral naturalness is the Mirror Twin Higgs model \cite{Chacko:2005pe}. The hidden sector in this model is an almost exact mirror copy of the visible sector. The only difference is the the vacuum expectation value (VEV) of the Higgs field, which also sets the fundamental fermion and electroweak gauge boson masses. The mirror Higgs VEV, denoted by $f$, is a factor of a few higher than the visible Higgs VEV $v$. In the minimal Mirror Twin Higgs model, this ratio $f / v$ is the only free parameter. This ratio is required to be $\gtrsim 3$ in order to suppress invisible decays of the Higgs to mirror sector particles via the Higgs portal, which are constrained to be below $10\%$ by LHC measurements~\cite{Burdman:2014zta}. 
On the other hand, $f/v \gtrsim 10$ makes the theory fine-tuned and hence unsuitable for addressing the Hierarchy Problem. 
Therefore, the most motivated Twin Higgs mirror sectors feature mirror quarks and leptons that are scaled up in mass by a sizeable $\mathcal{O}(1)$ factor relative to their SM cousins.

Besides the rather mild collider constraints on the Mirror Twin Higgs model, there are some interesting cosmological constraints on this setup. The Mirror Twin Higgs predicts a new massless gauge boson (the mirror photon) and three new light mirror neutrinos, all of which contribute to the total radiation density of the universe during Big Bang Nucleosynthesis and recombination. The number of light (relativistic) degrees of freedom (commonly denoted $N_\mathit{eff}$) at these times are cosmological observables that can be extracted from measured element abundances and from the cosmic microwave background. Crucially, the number of new light degrees of freedom $\Delta N_\mathit{eff}$ is constrained to be $\Delta N_\mathit{eff} \lesssim 0.25$ \cite{Aghanim:2018eyx}, or $\Delta N_{\mathit{eff}} \lesssim 0.49$ using the value of the Hubble constant obtained in \cite{Riess_2018}. The Mirror Twin Higgs predicts that the visible sector and the mirror sector should be in thermal equilibrium until the universe has cooled to temperatures of order $\mathcal O(\mathrm{GeV})$, which leads to the robustly excluded prediction $\Delta N_\mathit{eff} \approx 5.7$ \cite{Chacko:2016hvu, Craig:2016lyx}. There are various mechanisms that can avoid this issue -- most relevant for our purposes are those that generate a temperature asymmetry which effectively \emph{dilutes} the contribution to $\Delta N_\mathit{eff}$ from mirror sector particles. This can be achieved by asymmetric decays of some heavier particle (for instance, a right-handed neutrino), that preferentially decays to the mirror sector, heating up the mirror sector bath relative to the visible sector bath \cite{Chacko:2016hvu, Craig:2016lyx}. This can dilute the contribution to $\Delta N_\mathit{eff}$ and bring it within acceptable observational bounds. The resulting deviations are expected to be observable with next-generation measurements of the cosmic microwave background.

The Mirror Twin Higgs model is interesting in its simplicity and predictive power, but it is also a useful benchmark for more general models with complex dark sectors. Its similarity to the SM means that many calculations are very familiar, and it features a variety of interesting phenomena such as dissipative interactions (which allow mirror matter to lose energy and cool), bound states, and color confinement. Dissipative interactions in particular are interesting because they can lead to the formation of structure; in particular the formation of \emph{compact objects}
analogous to stars in the visible sector. In the Mirror Twin Higgs model it is very plausible that Mirror Stars could form \cite{Curtin:2019ngc, Curtin:2019lhm}, which can have their own distinctive (if extremely faint) electromagnetic signatures via the photon portal interaction. Furthermore, even if mirror stellar physics is radically different from SM stellar physics, the existence of degenerate remnant objects such as mirror white dwarfs and \emph{mirror neutron stars}, supported by degeneracy pressure, is a robust prediction. Their formation rate and initial mass function will depend on complex, non-linear mechanical and radiative feedback mechanisms, which are difficult to model even in the SM. 
Of course, the detection of even a single mirror neutron star would be a monumental discovery that would not only constitute a direct discovery of a component of dark matter, but also probe important details of the dark sector. The reach of current and future gravitational wave observatories make such a discovery a tantalising possibility.

It is important to stress that mirror neutron stars are not just a consequence of a single model of BSM physics; they are \emph{completely generic} predictions of any hidden sector model with dissipative self-interactions and some analogue of SM QCD. The fact that mirror sector matter is not charged under SM gauge groups (and hence does not directly produce any electromagnetic signatures) means that such objects could indeed be relatively common in our galaxy without having been detected thus far -- indeed their signatures are expected to be purely gravitational.

\subsection{Benchmark model}

We summarize here the specific details of the Twin Higgs mirror matter model we will use as a benchmark in our study of mirror neutron stars. For our purposes it consists of a dark QCD sector, presumed identical to that of the SM except that all of the quarks are a factor of $f/v$ heavier than the SM quarks. We will explicitly take values of $f/v$ in the motivated range $\sim 3 - 7$.
In order for the charge neutrality in nuclear matter to be satisfied, we also include the mirror electron and mirror muon along with mirror electromagnetic interactions. 
The mirror electron and muon are again a factor of $f/v$ heavier than their SM counterparts, and the strength of the mirror electromagnetic interaction is assumed to be unchanged, i.e. $\alpha' = \alpha$.
Mirror weak interactions proceed identically as in the SM, except that the mirror $W$ and $Z$ boson masses are also scaled up by the $f/v$ factor. However, the details of mirror weak interactions turn out to not be important for our study (it only matters that they are fast enough to maintain beta equilibrium inside the mirror neutron star). 

The heavier mirror quarks have two important consequences for the mirror QCD sector. The most important is that the increased masses of the heavy quarks -- specifically those heavier than $\Lambda_{QCD}$ -- will affect the running of the mirror strong coupling constant. This results in a  different \emph{confinement scale} $\Lambda_{QCD}'$ in the mirror sector. Following Ref.~\cite{Chacko:2018vss} we take the mirror confinement scale to be given by:
\begin{equation}
\label{eq:lambda_qcd_scaling}
    \frac{\Lambda_{QCD}'}{\Lambda_{QCD}} \approx 0.68 + 0.41 \log\left(1.32 + \frac{f}{v}\right),
\end{equation}
which is the result of a numerical fit to the renormalization group running of the mirror QCD coupling at 1-loop. For $f/v = 3 (7)$, $\Lambda_\mathrm{QCD}'/\Lambda_\mathrm{QCD} \approx 1.3 (1.6)$.
Another important effect is that the  mass of the mirror pion \emph{relative to the reference scale} $\Lambda_{QCD}'$ is also increased, since it scales as
\begin{equation}
\label{eq:mpi_scaling}
    m_\pi' \propto \sqrt{m_q' \Lambda_{QCD}'}.
\end{equation}

This essentially defines our mirror QCD sector, with one free parameter $f/v$. One can imagine this benchmark as a slice through a much broader space of QCD-like models, in which the masses of the quarks are free to vary independently rather than increasing in the same ratio.

There exists in the literature copious amounts of lattice data at large pion masses, which are often extrapolated down to SM values of the pion mass to obtain realistic results. However, for our purposes this data represents \emph{direct information} about the nature of the mirror QCD sector. We remind the reader that all quantities from lattice data are technically dimensionless. Thus the relevant parameter for us is the ratio $m_\pi'^2 / \Lambda_{QCD}'^2$, i.e. the size of the pion mass \emph{relative} to the reference scale.
In Section B we explain how this lattice data can be used to determine the mirror neutron star Equation of State.

\subsection{Comparison to previous studies}

The study of compact astrophysical objects made up of exotic matter has a rich history, but our work is the first to study the gravitational wave observables of \emph{realistic} mirror neutron stars that are made \emph{entirely} of invisible mirror matter \emph{arising from a dark QCD sector}, with  \emph{different properties than SM neutron stars} (focusing on the Twin Higgs mirror matter benchmark model described above).

There are of course many different kinds of exotic compact objects that could exist (see~\cite{Cardoso:2019rvt} for a review), including axion stars~(e.g. \cite{Eby:2015hsq, Clough:2018exo}) and Proca stars that can mimic black holes in gravitational wave events~\cite{CalderonBustillo:2020srq}. However, their properties are very unlike (mirror) neutron stars and they are made of entirely different types of matter, or arise in modifications of general relativity.

Neutron stars have been studied extensively as a probe of dark sectors in general and dark matter in particular. For example, dark matter captured inside (neutron) stars can affect their structure, which can lead to observable consequences~\cite{Goldman:1989nd,Berezhiani:2005vv,
Bertone:2007ae, Kouvaris:2010jy, deLavallaz:2010wp, Kouvaris:2011fi, McDermott:2011jp, 
Leung:2011zz, Kain:2021hpk, Bramante:2013hn, Bell:2013xk, Bertoni:2013bsa, Raj:2017wrv, Baryakhtar:2017dbj, Bell:2018pkk, Ellis:2018bkr, Garani:2018kkd, Dasgupta:2020dik, Garani:2020wge, Spolyar:2007qv,Goldman:2013qla,Goldman:2019dbq,Michaely:2019xpz,Ciancarella:2020msu,Berezhiani:2020zck} or even lead to entirely new classes of hybrid objects~\cite{Deliyergiyev:2019vti,Tolos:2015qra}; 
neutron stars can emit axions, leading to constraints from stellar cooling~\cite{Sedrakian:2015krq};
neutron stars could capture axion dark matter which gets converted to photons in its strong magnetic field~\cite{Hook:2018iia};
and phase transitions inside neutron stars have been studied as a possible probe of vacuum energy~\cite{Csaki:2018fls}. %

This is to be distinguished from our study of mirror neutron stars made entirely of mirror matter. 
Mirror neutron stars do not impact the properties or existence of SM neutron stars. While the mirror matter hypothesis predicts both to co-exist in our galaxy, their precise abundance depends on many aspects of the complete model.
Furthermore, since minimal mirror matter does not have astrophysically relevant interactions with SM matter (apart from gravity), and since its self-interactions give rise to different macroscopic parameters (like Jeans lengths and ionization temperatures) from SM matter, we do not expect most SM (mirror) neutron stars to form in regions with high concentrations of mirror (SM) matter, and hence do not expect them to necessarily contain large amounts of mirror (SM) matter.  We therefore focus on mirror neutron stars that are composed \textit{entirely} of mirror matter and are invisible to direct electromagnetic observations. 
That being said, the effects of $\mathbb{Z}_2$-symmetric mirror matter (i.e. exact SM copy) captured inside neutron stars has been studied  in~\cite{Sandin:2008db,Ciarcelluti:2010ji,Goldman:2011aa}, but this did not examine the possibility of mirror neutron stars made of mirror matter, nor did it consider the more general Twin Higgs mirror matter model.

In the context of our work we want to especially highlight the following previous studies. 
First, Ref~\cite{Beradze:2019yyp} did consider the possibility of mirror neutron stars and that they could make up a large fraction of observed gravitational wave events that would be observed in LIGO/Virgo. However, since this only considered the special case of fully $\mathbb{Z}_2$-symmetric mirror matter, the properties of mirror neutron stars would be identical to regular SM neutron stars, and the physics of mirror neutron stars themselves was therefore not the focus of the investigation. As we discussed in the main text, the possibility of an exact copy of SM matter as mirror matter is certainly an interesting possibility, but this is not motivated by the Hierarchy Problem, and represents only a single very special case of all the possibilities for mirror matter.
Our study is the first to realistically study the physics of mirror neutron stars within the Twin Higgs mirror matter model with significantly different properties than SM neutron stars. 

Second, Ref~\cite{Maedan:2019mgz} examines the possibility of a neutron star made of hidden sector nucleons arising as part of a dark-QCD sector, while~\cite{Narain:2006kx, Kouvaris:2015rea, Gresham:2018rqo} studied degenerate dark stars made of fundamental asymmetric dark matter fermions. This is much closer to the spirit of our investigation, and these studies explore how the mass-radius relationship of such dark neutron stars depend on some parameters of either a simplified dark hadron model, or simple interactions of the asymmetric dark matter.  
However, these calculations are extremely simplified compared to the motivated case of Twin  Higgs nuclear matter we want to study. In fact, to study degenerate stellar objects arising in any realistic model of dark QCD in this era of precision gravitational wave observations, one needs to take into account the effect of the several nucleon species; the exchange of and interactions amongst the various mesons; and the effect of the crust. 
We are able to take all these effects into account for the first time by drawing upon lattice  data for different  QCD parameters, and derive concrete predictions for gravitational wave and asymmetric pulsar binary observations. This is possible because our approach to Dark Complexity is guided by possible solutions to the hierarchy problem and its observable consequences, leading us to study Twin  Higgs nuclear matter in a regime that is different from the SM nuclear matter but still tractable to make robust and realistic predictions.

%
Exotic strongly interacting sectors with an ultra-light confining scale can give rise to even more exotic neutron star-like objects, see \cite{Alexander:2020wpm}. 
It is also 
important to keep in mind that in general, the properties of dark QCD sectors might be gravitationally probed by detecting the stochastic gravitational wave background produced during phase transitions to the confining phase, see  e.g. \cite{Schwaller:2015tja, Huang:2020mso}. Furthermore, the importance of gravitational wave measurements to probe exotic compact in general, whether they are fermion stars, boson stars, or other types, has been discussed in~\cite{Giudice:2016zpa}.

Finally, Ref~\cite{Dvali:2019ewm} represents an investigation of Dark Complexity that is very close to ours philosophically in that it takes guidance from other fundamental puzzles in high energy physics, but very different in the kind of exotic stellar object it predicts. This study takes inspiration from theories that solve  the hierarchy problem by predicting large-$N$ number of copies of SM, each of which are incredibly dilute and only interact with each other gravitationally. This predicts microscopic exotic objects that contain material from all $N$ SM-like sectors. 
This is very different from mirror neutron stars, but is an interesting astrophysical consequence of another potential solution to the hierarchy problem.

\section{Mirror Neutron Star Equation of State}
%
\label{sec:EoS}

We first build a reasonable model for the equation of state (EoS) of SM neutron stars, which can be rescaled to construct a mirror neutron star EoS. 
Neutron stars probe different degrees of freedom across their radius, or equivalently their baryon density $n_B$.
The uncertainty is largest at the core of a neutron star, where $n_B\gtrsim 2n_B^{sat}$, with $n_B^{sat} \approx 3 \times 10^{14} g/cm^3$ being the nucleon saturation density \cite{Baym:2017whm}.
Starting from their crust and ending at their core, their  degrees of freedom are thought to be: atoms, then nuclei (with increasing mass), nuclei and free nucleons (after the neutron drip line), nucleonic pasta, interacting nucleons (above $n_B^{sat}$), then possibly hyperons, and possibly quarks (though we do not consider these hypothetical phases here).

Due to the sign problem \cite{Troyer:2004ge}, one cannot directly calculate the 
EoS of a neutron star from QCD. Effective models are used instead.  
Nuclear physicists rely on a rich trove of data from  nuclear experiments, the masses of known nuclei, measurements of nucleonic interactions, Lattice QCD calculations of resonances and interactions, and chiral effective theory to define reasonable EoSs up to  $ n_B^{sat}$ \cite{Tews:2018iwm}. A well-established model used here
is the  Baym-Pethick-Sutherland crust~\cite{Baym:1971pw}, which uses free nuclei and electrons. Given $n_B$, one finds the nucleus that minimizes the energy, modeling nuclei from either known experimental measurements~\cite{reference.wolfram_2020_isotopedata} or extrapolating from the semi-empirical mass formula \cite{Weizscker:1935zz}.  The EoS is then constructed from well-known thermodynamic relations.

We model the core using a mean-field approach that incorporates electrons, muons, neutrons, and protons in the zero-temperature approximation valid for $T \ll \Lambda_{QCD}$. 
Interactions are mediated via the mesons $\sigma$, $\omega^\mu$, $\vec{\rho}^{\mu}$, which may be attractive, repulsive, or self-interactions among mesons. We ensure local charge neutrality and chemical equilibrium under weak interactions. The model contains 12 parameters,  fixed by experimental data when possible, determined from well-known scaling relationships, or constrained by nuclear experimental or astrophysical neutron star data~\cite{Chen:2014sca,Baym:2017whm,Dexheimer:2018dhb}.

We connect the crust and the core EoS, from the liquid gas phase transition density to the saturation density, through an interpolation function.
This interpolation should not significantly affect our results, as details of the inner/outer crust have been estimated to affect low-mass neutron stars at the sub-10\% level~\cite{Baym:2017whm}.

Solving for the structure of SM neutron stars with our EoS yields reasonable results and passes typical physics benchmarks, 
such as  sensible values for $n_{sat}$, binding energy per nucleon, incompressibility, and symmetry energy.  It fits within all known constraints from NICER \cite{Riley:2019yda,Miller:2019cac} and LIGO/VIRGO \cite{TheLIGOScientific:2017qsa,Abbott:2018exr} for the radius (see Fig.\ \ref{fig:MRplotfront}) and tidal constraints (see Fig.\ \ref{fig:mirrorobservablesplot}). 
Our SM EoS leads to a maximum mass of $M_{max}\sim2.1 \pm 0.1 M_\odot$, consistent with millisecond pulsars~\cite{Cromartie:2019kug}.  

With our EoS we can now smoothly rescale several of its parameters to derive a mirror neutron star EoS. 
This rescaling accounts for different masses and couplings in mirror matter due to mirror quarks/leptons being heavier than SM fermions by a factor  $f/v \sim 3 - 10$.

Particle masses and nucleon-meson couplings rely on knowledge from low energy hadron physics \cite{GellMann:1960np}, first principle Lattice QCD (e.g. \cite{Bratt:2010jn}),  
and chiral perturbation theory \cite{Hanhart:2008mx,Molina:2020qpw}.
Our results are robust to variations in the scalings of these parameters within uncertainties from lattice QCD or chiral perturbation theory.
All assumptions, parameters, error bars, scalings,  constraints, and plots of the EoS, are collected in Tab. \ref{tab:params}-\ref{tab:constraints} and Fig.\ \ref{fig:EoS}.
Future lattice QCD investigations could improve our calculations by supplying information on the $m_\pi$ dependence of quartic meson interactions.

Our final EoS are shown in Fig.~\ref{fig:EoS}, plotting the pressure $p$ versus energy density  $\epsilon$ for several different values of the quark mass rescaling $m_{q'}/m_q = f/v$. Shaded regions represent uncertainties discussed in Sec.~\ref{sec:scalinguncertainty}.


\subsection{Core Model for a SM Neutron Star}

Inside the core of a neutron star, the strong compression leads to a very large density of (degenerate) neutrons. 
Within neutron stars, the low-momentum electron energy levels are filled, which prevents $\beta$ decay $n\rightarrow p+e^-+\bar{\nu}_e$; therefore, the weak interaction of inverse $\beta$ decay dominates (wherein protons are converted to neutrons), with the neutrinos produced in this process free to escape the star.  Chemical equilibrium under weak interaction, thus, leads to a very neutron rich environment.  Then, the remaining protons in the neutron star must be balanced by an exact equal number of leptons in order to ensure charge neutrality. At sufficiently high densities, electrons can also convert into muons through the reaction $e^- \to \mu^- + \bar \nu_\mu + \nu_e$,
which will thus be present in equilibrium. Hence, weak interactions must enter the core EoS through chemical equilibrium conditions \cite{glendenningbook}. On the other hand, compared to strong interactions, weak couplings provide only negligible corrections to the EoS. Therefore, our core model does not need to account for the details of weak interactions, as long as those are fast enough to maintain beta-equilibrium but feeble enough to not generate significant interparticle potentials. Our core model thus consists of strongly interacting protons and neutrons, in the presence of a free gas of electrons and muons, all in chemical equilibrium.

The EoS of  strongly interacting matter at very high densities is not amenable to first principle QCD calculations and must be extracted from an effective model.  We employ a simple relativistic mean-field model including effects from scalar attraction, vector repulsion and vector-isovector interactions between protons and neutrons, mediated by the mesons $\sigma$, $\omega^\mu$ and $\vec\rho^{\;\mu}$\cite{Walecka:1974qa,glendenningbook,Horowitz:2000xj,Chen:2014sca,Dexheimer:2018dhb,Fattoyev:2020cws}. 
In such models, the balance between attractive scalar interactions and repulsive forces mediated by vector mesons leads to \emph{nuclear saturation} --- the existence of an optimal density at which the energy per baryon reaches its minimum. Vector-isovector interactions, on the other hand, are responsible for increasing the energy cost of the excess of neutrons over protons. 
For a better description of nuclear matter and astrophysical constraints, we include self-interactions between sigma mesons and interactions between vector and vector-isovector mesons, respectively \cite{glendenningbook,Dexheimer:2018dhb}. 
The small difference between the neutron and proton masses is neglected, making the core model isospin symmetric. (The neutron-proton mass difference in Twin Higgs mirror matter is larger than in the SM~\cite{Chacko:2018vss}, but still small enough to be neglected for our purposes.) However, this symmetry is broken by a finite isospin chemical potential, generated by the different electric charges of neutrons/neutrinos and protons/electrons. 
The hadronic part of our core model consists of the following Lagrangian:
\begin{eqnarray}
\label{eq:LagNucMes}
\mathcal{L} &=&
\bar\psi\big[i \, \gamma_\mu \partial^\mu - m_B + \gamma^0\,\mu_B + \gamma_0\,\dfrac{\tau_3}{2}\,\mu_I\big]\psi +
\\  \nonumber  && 
+\bar\psi\big[
- g_\omega\, \omega^\mu \,\gamma_\mu -g_\rho\, \gamma^\mu\,\vec\rho_\mu \cdot \dfrac{\vec\tau}{2} + g_\sigma \, \sigma \big]\psi 
\\  \nonumber  && 
+ \dfrac{1}{2} \partial_\mu\sigma\partial^\mu\sigma  
-  \dfrac{1}{4} \omega^{\mu\nu}\omega_{\mu\nu} 
- \dfrac{1}{4}\,\vec\rho_{\mu\nu}\cdot \vec \rho^{\mu\nu} + 
 \\  \nonumber  && 
-\dfrac{1}{2}m_\sigma^2\,\sigma^2 
+ \dfrac{1}{2} m_\omega^2 \, \omega^\mu \omega_\mu 
+\dfrac{1}{2}\,m_\rho^2\,\rho^\mu\rho_\mu  
 \\  \nonumber  && 
 - \dfrac{a_3}{3}\, m_B\, (g_\sigma \,\sigma)^3 - \dfrac{a_4}{4}\, (g_\sigma\, \sigma)^4
+ g_{\omega\rho}\,(g_\omega\,\omega^\mu)^2(g_\rho\,\vec\rho^\mu)^2 \,,
\end{eqnarray}
where $m_B$ is the vacuum nucleon mass, $\mu_B$ is the baryon chemical potential, $\mu_I$ is the isospin chemical potential, $\omega_{\mu\nu}\equiv \partial_\mu\omega_\nu-\partial_\nu\omega_\mu$ and  $\vec\rho_{\mu\nu} \equiv \partial_\mu \vec \rho_\nu -  \partial_\nu\vec \rho_\mu
- g_\rho\, (\vec\rho_\mu \times \vec\rho_\nu)$.
(Note that the chemical potentials are derived as a function of baryon density below.)
Here, we employ natural units $\hbar=c=1$, where $c$ is the speed of light, and the metric signature $\eta_{\mu\nu} = \operatorname{diag}\, (1,-1,-1,-1)$.
The nucleon field $\psi = (p, n)^T$ in Eq.~\eqref{eq:LagNucMes} has isospin indices corresponding to the proton and neutron degrees of freedom, with isospin given by $\vec I = \frac{1}{2}\,\vec \tau$. 
The last three terms in Eq.~\eqref{eq:LagNucMes} are arranged so that the properties of infinite nuclear matter in equilibrium depend on the nucleon-meson couplings $g_\sigma$, $g_\omega$ and $g_\rho$ only through ratios $g_\sigma/m_\sigma$, $g_\omega/m_\omega$ and $g_\rho/\omega_\rho$.

We work in the mean-field approximation and assume low temperatures compared to typical Fermi energies: $T\approx 0$, valid if $T \ll \Lambda_\mathrm{QCD}$. In this approximation, we only consider the vacuum-expectation values acquired by fields: 
\begin{equation}
\label{eq:condensates}
\vev{ \sigma} \equiv \bar\sigma\,,\qquad 
\vev{\omega^\mu } \equiv \bar\omega \,\delta^\mu_0\,,\qquad 
\vev{\rho^\mu_i} \equiv \bar\rho \,\delta_{i 3}\,\delta^\mu_0\,,
\end{equation}
where we require that the symmetry under real-space rotations and the action of the isospin operator $I_3$ are preserved. 
Under this approximation, the nucleons behave as a free gas of fermions with 
effective mass and chemical potentials given by the mesonic condensates:
\begin{equation}
\label{eq:effparams}
m_B^* \equiv m_B - g_\sigma \bar \sigma\,,\quad
\mu^*_B\equiv \mu_B - g_\omega \,\bar\omega\,,\quad
\mu^*_I \equiv \mu_I - g_\rho \,\bar\rho\,.
\end{equation}

When calculating the EoS, we  consider the condensates in Eq.~\eqref{eq:condensates} to be uniform in space. In practice, they will change slowly with radius via their dependence on $n_B$ and $n_I$ \cite{glendenningbook}. 
We first find the EoS as a function of the baryon and isospin densities $n_B$ and $n_I$. 
The effective chemical potentials for the proton and the neutron can be found from the respective densities, $n_p = n_B/2+n_I$ and $n_n = n_B/2 - n_I$:
\begin{equation}
\mu_{p,n}^* = \sqrt{{k^{F}_{p,n}}^2 + m_B^{*2}}\,,\qquad 
 k^F_{p,n} =  2\,\pi\,\left( \dfrac{3}{2\times4\pi}\,n_{p,n} \right)^{1/3}\,,
\end{equation}
where $k^F_{p,n}$ is the Fermi momentum.
From them we find the effective baryonic and isospin chemical potentials
\begin{equation}
\label{eq:effchempot}
\mu_{B}^*= \dfrac{\mu_{p}^*+\mu_{n}^*}{2}\,,\qquad
\mu_{I}^* = \mu_{p}^*-\mu_{n}^*\,.
\end{equation}
The pressure can be found from the spatial components of the energy-momentum tensor:
\begin{equation}
    p = \langle \mathcal{L} \rangle + \frac{1}{3}\langle \bar\psi \,i\,\slashed{\partial}\,\psi\rangle\,.
\end{equation}
In practice, it is the pressure of a free gas of nucleons with the effective parameters in Eq.~\eqref{eq:effparams}, plus terms from the mesons: 
\begin{multline}
p = p_{\textrm{free}}(\mu_{p,n}^*,m_B^*) 
- \dfrac{1}{2}m_\sigma^2\,\bar\sigma^2  
+ \dfrac{1}{2}m_\omega^2 \bar\omega^2  
+ \dfrac{1}{2}m_\rho^2 \bar\rho^2 +\\
\;\,
-\dfrac{a_3}{3}\, m_B\, (g_\sigma \,\bar\sigma)^3 - \dfrac{a_4}{4}\, (g_\sigma\, \bar\sigma)^4
+ g_{\omega\rho}\,(g_\omega\,\bar\omega)^2(g_\rho\,\bar\rho)^2
\end{multline}
The energy density can be found from the time component of the energy-momentum tensor or, equivalently, from the first law of thermodynamics:
\begin{equation}
\epsilon = \mu_B\,n_B - \mu_I\, n_I   - p\,.
\end{equation}

The condensates can be found from Eq.~\eqref{eq:effparams} and the Euler-Lagrange equations:
\begin{equation}
\label{eq:stationary}
\frac{\partial \langle\mathcal{L}\rangle}{\partial \bar\sigma}  =  \frac{\partial \langle\mathcal{L}\rangle}{\partial \bar\omega} =  \frac{\partial \langle\mathcal{L}\rangle}{\partial \bar\rho} = 0\,.
\end{equation}
The condensates $\bar\omega$ and $\bar \rho$ are found as functions of $n_B$ and $n_I$,  and we can solve Eqs.~\eqref{eq:effparams} and \eqref{eq:effchempot} for $\mu_I$ and $\mu_B$. Solving Eq.~\eqref{eq:stationary} for $\bar\sigma$, we also find $m_B^*$. 

Finally, local charge neutrality requires that we also
include a density of negatively charged leptons 
$n_\ell$. Requiring local charge neutrality and 
beta equilibrium leads to:
\begin{equation}
n_\ell=n_p=n_B/2+n_I\,, \qquad \mu_I =\mu_p - \mu_n = - \mu_\ell\,,
\label{eq:weakequilibrium}
\end{equation}
with the same chemical potential $\mu_\ell$ for all lepton flavors. 
We include both electrons and muons, modeled as free fermions at zero temperature. 
Solving Eq.~\eqref{eq:weakequilibrium}, we determine the proton and lepton fractions in the core, as functions of $n_B$.

The core model contains 12 parameters: 6 masses and 6 couplings, described in Table~\ref{tab:params}. The SM value of the parameters are shown in the third column of this table. Our results are only sensitive to the combinations $g_\sigma/m_\sigma$, $g_\omega/m_\omega$ and $g_\rho/m_\rho$, rather than $g_\sigma$, $g_\omega$, $g_\rho$ and the corresponding masses separately. However, to make the rescalings for mirror matter more transparent, we fix masses and couplings separately. Mass parameters are fixed to central experimental values \cite{Zyla:2020zbs}, while couplings are chosen to reproduce properties of nuclear matter at saturation  \cite{Glendenning:1987gk,Glendenning:1987gd,Johnson:1987zza,Sharma:1988zza,Moller:1988bbe,Jaminon:1989wj,Li:1992zza,Li:1992zza,Kruger:2013kua} and neutron star observations \cite{Demorest:2010bx,Antoniadis:2013pzd,Abbott:2018exr,Cromartie:2019kug,Riley:2019yda}. The values we take for each of these constraints are shown in Table~\ref{tab:constraints}, where we also show the parameters for which each constraint is most relevant. Values for the effective nucleon mass, nuclear incompressibility and symmetry energy at saturation are chosen within acceptable ranges so as to satisfy neutron star physics constraints.  In Table~\ref{tab:constraints} we do not require that our EoS reaches an even larger maximum mass, as was measured in GW190814 \cite{Abbott:2020khf}, since it is still under debate if the secondary compact object in that event was a neutron star or a black hole \cite{Most:2020bba,Broadhurst:2020cvm,Tan:2020ics,Fishbach:2020ryj,Dexheimer:2020rlp,Tsokaros:2020hli,Fattoyev:2020cws}.

\begin{table*}[t]
\begin{center}
\begin{tabular}{|c|c|c|c|c|c|c|}
\hline
{\multirow{2}{*}{Particle}} & {\multirow{2}{*}{Parameters}} & {\multirow{2}{*}{SM Value}} &  \multicolumn{3}{c|}{Mirror pion mass scaling, Eqn~(\ref{mass_fit})} & {\multirow{2}{*}{Source}}\\ \cline{4-6}
 & &  & $b_0$ & $b_1$ & $b_2$ &  \\
\hline
\hline
$\pi$ & $f_\pi$ & $92.07\pm 0.85$ MeV \cite{Zyla:2020zbs} & $0.094_{-0.005}^{+0.005}$ & $0.067_{-0.011}^{+0.011}$ & $0.06_{-0.10}^{+0.10}$ & \cite{Bratt:2010jn,Horsley:2013ayv}\\
\hline
\hline
{\multirow{1}{*}{$n, p$} }
& $m_B$ &  $938.9 \pm 0.6$ MeV$^a$  \cite{Zyla:2020zbs}& $0.933_{-0.006}^{+0.003}$ & $   1.82_{-0.09}^{+0.12}$ & $-1.35_{-0.35}^{+0.25}$ &  \cite{WalkerLoud:2008bp,Syritsyn:2009mx,Bratt:2010jn,Jung:2012rz} \\
\hline
\hline
{\multirow{4}{*}{$\sigma$}} 
& $m_\sigma$ &  400--550 MeV \cite{Zyla:2020zbs} & $0.408_{-0.001}^{+0.012}$ & $2.42_{-0.07}^{+0.40}$ & $-11.0_{-6.6}^{+0.2}$ &  \cite{Hanhart:2008mx,Albaladejo:2012te,Pelaez:2015qba} \\
\cline{2-7}
& $g_\sigma$ &  7.95$^b$
& \multicolumn{3}{c|}{ $g_\sigma \propto m_B/f_\pi$$^c$
}  
& \cite{Goldberger:1958tr,GellMann:1960np,Koch:1997ei} \\
\cline{2-7}
& $a_3$ & 0.0036$^b$
&\multicolumn{3}{c|}{Kept constant.$^d$
}& --- \\
\cline{2-7}
& $a_4$ &  0.0059$^b$
&\multicolumn{3}{c|}{Kept constant.$^d$
} &  ---\\
\hline
\hline
{\multirow{2}{*}{$\omega$}}
& $m_\omega$ &  $782.65\pm 0.12$ MeV  \cite{Zyla:2020zbs}&  $0.773_{-0.010}^{+0.015}$ & $0.573_{-0.028}^{+0.120}$ & $0.659_{-0.411}^{+0.004}$ & \cite{McNeile:2009mx} \\ 
\cline{2-7}
& $g_\omega$  & 9.23$^b$
& \multicolumn{3}{c|}{$g_\omega \propto g_\rho$} & --- \\
\hline
\hline
{\multirow{3}{*}{$\rho$} }
& $m_\rho$  & $775.26\pm 0.25$ MeV \cite{Zyla:2020zbs} & $0.746_{-0.013}^{+0.017}$ & $0.659_{-0.028}^{+0.120}$ & $0.653_{-0.391}^{+0.007}$ &  \cite{Molina:2020qpw} \\
\cline{2-7}
& $g_\rho$  & 39.19$^b$
& \multicolumn{3}{c|} {$g_\rho \propto m_\rho/(\sqrt{2}\,f_\pi)$$^e$
} &  \cite{Riazuddin:1966sw,Birse:1996hd,Molina:2020qpw} \\
\cline{2-7}
& $g_{\omega\rho}$  & 0.2$^b$
&\multicolumn{3}{c|}{Kept constant.$^d$
} & ---\\
\hline
\hline
$e^-$ & $m_e$  & 0.511
MeV$^f$  \cite{Zyla:2020zbs} & \multicolumn{3}{c|}{ $m_e\propto m_q$} &---\\
\hline
\hline
$\mu^-$ & $m_\mu$ & 105.658
MeV$^f$ \cite{Zyla:2020zbs}& \multicolumn{3}{c|}{$m_\mu\propto m_q$} &  ---\\
\hline
\end{tabular}
\end{center}
    \caption{
    Parameters of our core model, grouped by their corresponding particles. Their Standard Model 
    (SM) values were obtained from the constraints in Table~\ref{tab:constraints} and from the Particle Data Group \cite{Zyla:2020zbs}. The quark-mass dependence used to rescale the parameters in the Mirror Sector (MS) are extracted from Lattice-QCD data and phenomenological relations, when these are available.  
    We note that the meson couplings and masses are only relevant to the physics in equilibrium through combinations of the form $g/m$. 
    See footnotes below for additional information on individual parameters.
    \\
    \footnotesize\textsuperscript{a} 
    Uncertainty due to the difference between the neutron and proton masses.
    \\
    \footnotesize\textsuperscript{b} 
    Obtained from the constraints on Table~\ref{tab:constraints}.
    \\
    \footnotesize\textsuperscript{c}
    From the Goldberger-Treiman relation (GTR), the pion-nucleon coupling is $g_{\pi} = { g_A}\,m_B/f_\pi$, and from chiral symmetry $g_\sigma \simeq g_\pi$ \cite{Goldberger:1958tr,GellMann:1960np,Koch:1997ei}. 
    \\
    \footnotesize\textsuperscript{d}
    In the Lagrangian Eq.~\eqref{eq:LagNucMes}, these couplings are already rescaled with the appropriate powers of other parameters.
    \\
    \footnotesize\textsuperscript{e}
     From the KSFR relation $g_\rho=m_\rho/(\sqrt{2}\,f_\pi)$ \cite{Riazuddin:1966sw,Birse:1996hd,Molina:2020qpw}.
    \\
    \footnotesize\textsuperscript{f}
     Uncertainties are vanishingly small for our purposes.
    }  
    \label{tab:params}
\end{table*}

\begin{table*}[]
    \centering
\begin{tabular}{|c|c|c|c|}
\hline
Constraint & Description & SM Value & Parameters \\
\hline
$n_B^{\textrm{sat}}$ & Nuclear saturation density &  0.153 fm$^{-3}$ 
& \multirow{2}{*}{$g_\omega$} \\
$E_b$ & Binding energy per nucleon & -16.3 MeV  & \\
\hline
$m_B^{* \textrm{sat}}$  & Nucleon (Dirac) mass at saturation & 0.76 $m_B$ 
& \multirow{2}{*}{$g_\sigma$, $a_3$, $a_4$}\\
$K^{\textrm{sat}}$ & Nuclear incompressibility at saturation & 300 MeV 
& \\
\hline
$a^{\textrm{sat}}_{\textrm{sym}}$ & Symmetry energy at saturation & 28 MeV 
& $g_\rho$\\
\hline
$R$ & Star radii constraints & \cite{Abbott:2018exr,Riley:2019yda} & $g_{\omega\rho}$\\ 
$M_\textrm{max}$ & Maximum neutron star mass & $M_\textrm{max}\gtrsim 2\,M_\odot$ 
& --- \\ 
\hline
\end{tabular}
    \caption{Nuclear physics \cite{Glendenning:1987gk,Glendenning:1987gd,Johnson:1987zza,Sharma:1988zza,Moller:1988bbe,Jaminon:1989wj,Li:1992zza,Li:1992zza,Kruger:2013kua} and astrophysics \cite{Demorest:2010bx,Antoniadis:2013pzd,Abbott:2018exr,Cromartie:2019kug,Riley:2019yda} constraints used to fix the parameters of the model for the neutron star EoS, shown in Table~\ref{tab:params}. Values for the effective nucleon mass, nuclear incompressibility and symmetry energy at saturation are optimized to satisfy constraints on neutron star masses and radii.}
    \label{tab:constraints}
\end{table*}

\subsection{Core Model for a Mirror Neutron Star}

To extend our core model to the mirror sector, we need to understand how its parameters scale with the mirror Higgs VEV, through the quark masses $m_q$. Parameters of the hadronic model will depend on $m_q'/m_q$ indirectly, through $\Lambda_\mathrm{QCD}$ and $m_\pi$, which scale  according to Eqs.~\eqref{eq:lambda_qcd_scaling} and \eqref{eq:mpi_scaling}. Lepton masses are scaled in proportion to the Higgs VEV and are thus $\propto m_q'/m_q$. Since weak couplings are neglected in the core EoS, no weak-interaction parameters need to be rescaled, as long as electron capture and inverse beta decays  are sufficiently fast to maintain beta equilibrium. Moreover, because the isospin asymmetry is driven by beta equilibrium, we neglect the mass difference between neutron and proton in our core model, eliminating the need to rescale this quantity as well.

The extension of our core model to mirror QCD primarily involves scaling the masses and couplings that appear in Eq.~\eqref{eq:LagNucMes} as a function of $\Lambda_{QCD}'$ and $m_\pi'$. In general, we can express the mass of a given resonance as a power series in $m_\pi^2/\Lambda_{QCD}^2$ as follows:
    \begin{multline}
    \label{mass_fit}
   \frac{m'_i/\mathrm{GeV}}{\Lambda'_{QCD}/\Lambda_{QCD}} = b_0 + b_1 \left( \frac{m'_\pi/\mathrm{GeV}}{\Lambda'_{QCD}/\Lambda_{QCD}}\right)^2+\\
   +b_2 \left( \frac{m'_\pi/\mathrm{GeV}}{\Lambda'_{QCD}/\Lambda_{QCD}}\right)^4 + \dots\,,
    \end{multline}
    where the coefficients $b_0$, $b_1$ and $b_2$ are extracted, up to quadratic order, from lattice data and Chiral Perturbation Theory with dispersion theory (see Table~\ref{tab:params} for details and references). The pion decay constant $f_\pi$ is rescaled in the same way. Although it does not appear explicitly in our Lagrangian, it is used to estimate the dimensionless couplings $g_\rho$, $g_\omega$, $g_\sigma$, as explained below. The lepton masses $m_e$ and $m_\mu$, on the other hand, are taken to scale as the quark masses, proportionally to the Higgs VEV. Errors quoted on the scaling parameters in Table~\ref{tab:params} reflect uncertainties presented in the respective references, and in cases where we have consulted multiple references we use uncertainties that reflect the spread of the collective data, if larger than the uncertainties in individual references.
    
The remaining, dimensionless parameters that appear in our Lagrangian, on the other hand, are not extracted from lattice data. The meson-nucleon couplings $g_\sigma$, $g_\omega$ and $g_\rho$ are instead obtained via relations from low-energy hadronic physics. The $\rho$-nucleon coupling constant, for instance, is related to $m_\rho/f_\pi$ by the so-called KSFR relation $g_\rho=m_\rho/(\sqrt{2}\,f_\pi)$ \cite{Riazuddin:1966sw,Birse:1996hd}, the accuracy of which is discussed in \cite{Molina:2020qpw}. The $\omega$-nucleon coupling is taken to rescale similarly: $g_\omega/m_\omega \propto g_\rho/m_\rho$. 
    Moreover, the scalar-nucleon coupling $g_\sigma$ can be related to the nucleon mass through the Goldberger-Treiman relation, $g_\sigma = g_{\pi NN} = g_A\,M_N/f_\pi$, where $g_A\sim \mathcal{O}(1)$ is the axial charge of the nucleon and, inspired by chiral symmetry, we assume that the sigma-nucleon and pion-nucleon couplings are approximately equal 
    \cite{Goldberger:1958tr,GellMann:1960np,Koch:1997ei}. Since $g_A$ is approximately constant as a function of the quark mass, we take  $g_\sigma \propto M_N/f_\pi$.

Finally, for simplicity and lack of guidance, the remaining coupling constants $a_3$, $a_4$ and $g_{\omega \rho}$ are not rescaled with $\Lambda_{QCD}$ and $m_\pi$. 
Notice that the full meson-meson coupling constants in Eq.~\eqref{eq:LagNucMes} are actually $a_3 g_\sigma^3 m_B/3$, $a_4 g_\sigma^4/4$ and $g_{\omega\rho}g_\omega^2g_\rho^2$. 
They are written in such a way that the physics of infinite, homogeneous nuclear matter depends on the nucleon meson couplings only through the ratios $g_\sigma/m_\sigma$, $g_\omega/m_\omega$ and $g_\rho/m_\rho$. 
Because $a_3$, $a_4$ and $g_{\omega \rho}$ are small and dimensionless, it is reasonable to assume that the full couplings will depend on $m_q'/m_q$ mainly through the rescaling of the larger parameters $g_{\sigma,\omega,\rho}$. 
To make sure our conclusions are robust, despite that assumption, we have checked the effect of changing $a_3$, $a_4$ and $g_{\omega \rho}$ in both the SM and mirror sector scenarios. We find in particular that neutron star properties are most sensitive to the coupling $g_{\omega\rho}$ (recall that $a_3$ and $a_4$ are chosen to satisfy nuclear physics constraints, see Table~\ref{tab:constraints}). While the value of the maximum mass does not depend sensitively on $g_{\omega\rho}$, we find that the neutron star radius constraints from NICER~\cite{Riley:2019yda,Miller:2019cac} and LIGO/VIRGO~\cite{TheLIGOScientific:2017qsa,Abbott:2018exr,Abbott:2018wiz} are only satisfied for $g_{\omega\rho}$ in the range $0.1 \lesssim g_{\omega\rho} \lesssim 0.25 $ (smaller values of $g_{\omega\rho}$ correspond to larger radii, and vice versa). While changing the values of these couplings can significantly impact mass-radius relations and properties of neutron star matter, we have found that it did not affect the overall scaling of the mass-radius curve with $m_q'/m_q$ (though this conclusion could be modified if $a_3, a_4, g_{\omega \rho}$ themselves have a strong independent dependence on $m_\pi/\Lambda_\mathrm{QCD}$).

The lack of data for the $a_3, a_4, g_{\omega \rho}$ couplings as functions of the quark masses (or equivalently $m_\pi/\Lambda_\mathrm{QCD}$) is a limitation to our mirror neutron star model which leaves room for improvement. It would be very interesting if future lattice calculations could extract the pion mass dependence of these dimensionless quartic meson couplings. 

\begin{figure}
    \centering
    \includegraphics[scale=0.4]{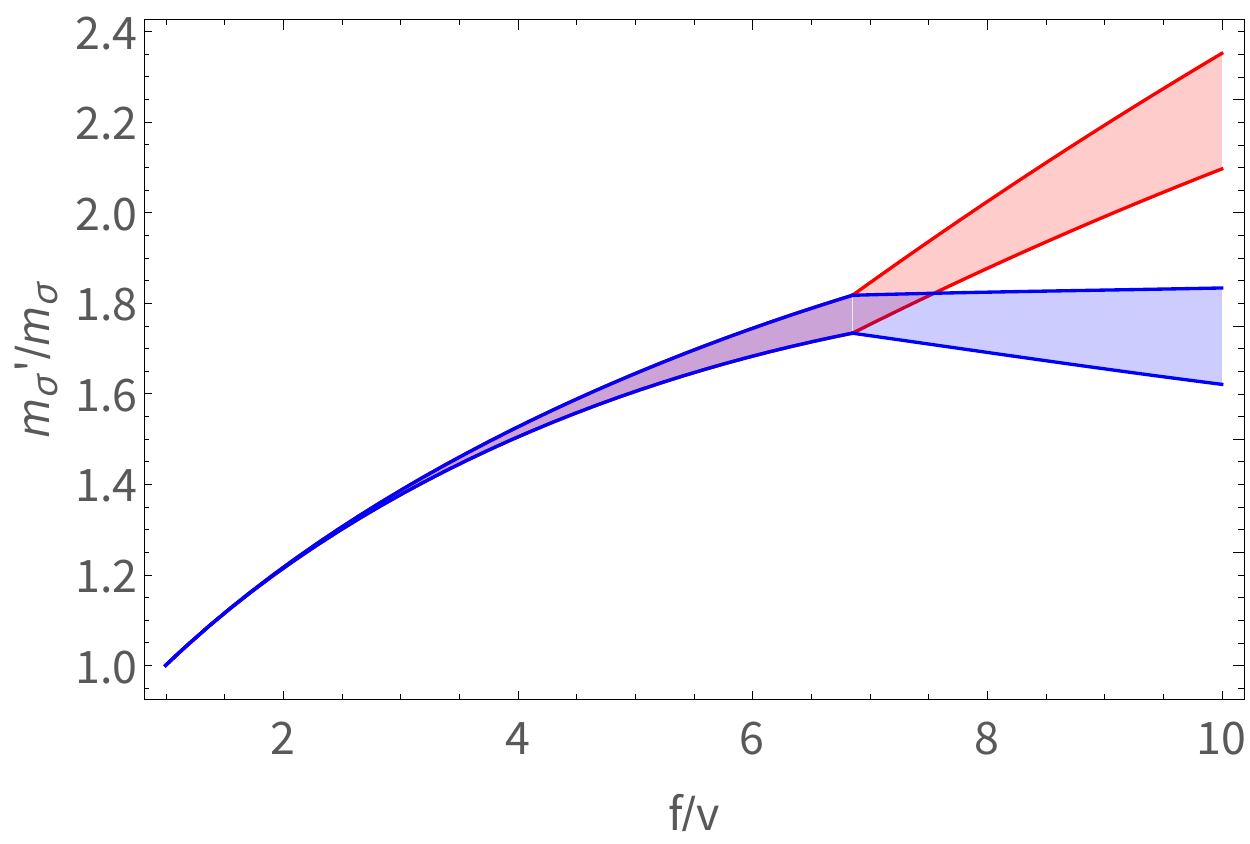}
    \caption{Mass of the sigma meson as a function of $f/v$, employing data from  Chiral Perturbation Theory with dispersion theory Ref.~\cite{Hanhart:2008mx,Pelaez:2015qba}. The bands correspond to the uncertainty due to different lattice fits. Note in particular the splitting at around $f/v \approx 7$. }
    \label{fig:sigma_mass}
\end{figure}

The procedure outlined above is complicated somewhat by the peculiar behaviour of the sigma mass at large pion masses. As discussed in Ref.~\cite{Pelaez:2015qba}, the dependence of the $m_\sigma$-pole on $m_\pi$ appears to split into two distinct trajectories at around $m_\pi \approx 300$ MeV. This may indicate that at this pion mass the sigma is replaced by two distinct resonances, whose mass splitting increases with increasing pion mass. 
We do not attempt to model this possibility, instead restricting our focus to the model outlined above, with just one sigma-like resonance. We will therefore present our results for each of the two $m_\sigma$ trajectories. It is worth noting however that $m_\pi \approx 300$ MeV corresponds to $f/v \approx 7$, so the splitting issue does not actually manifest until relatively high values of $f/v$. In Table~\ref{tab:params} we present only the scaling coefficients for the sigma mass dependence before the bifurcation occurs, since this regime indeed will dominate the results we present in this study.
In Figure~\ref{fig:sigma_mass} we show this splitting as a function of $f/v$, including our uncertainty bands derived from the spread of different lattice fits presented in Ref.~\cite{Pelaez:2015qba}. Regardless if one includes the lower mass or the higher mass of $\sigma$ for $f/v \gtrsim 7$, one still always sees that the mass and radius shrink with increasing values of $f/v$.  However, the degree to which the mass and radius inverse scale with $f/v$ \emph{does} depend on the bifurcation.

\subsection{Crust model for a SM Neutron Star}

The EoS for the neutron star crust is constructed following the procedure outlined in \cite{Baym:1971pw}. We treat the crust as being composed of nuclei and free electrons. Because of their higher mass, muons are not present at crust densities, where $\mu_e<m_\mu$. At a given baryon density/depth we can assume that all nuclei are of the single species which 
minimizes the total energy per unit volume:
\begin{equation}
    E_{tot} = n_N (W_N + W_L) + E_e,
\end{equation}
where $n_N$ is the number density of nuclei, $W_N$ is the energy of an isolated nucleus, $W_L$ is the electrostatic lattice energy per nucleus, and $E_e$ is the total electron energy per unit volume. The nuclear energy is given by 
\begin{equation}
 W_N = m_n\, (A - Z) +m_p \,Z - b_N\,A,
\end{equation}
where $b_N$ is the binding energy per nucleon for the relevant species.
The lattice energy is given by~\cite{1960JMP.....1..395C}
\begin{equation}
\label{e.WL}
 W_L = - \kappa \frac{Z^2 e^2}{4\pi\epsilon_0 a},
\end{equation}
where 
for a body-centred cubic lattice, which is the configuration which minimizes the lattice energy \cite{Baym:1971pw, 1960JMP.....1..395C}, $\kappa = 1.8119620$ and the lattice constant $a$  is related to $n_N$ via $n_N a^3 = 2$.

Treating the electrons as free, their energy is given by
\begin{align}
\label{e.crustEe}
E_e &= \int_0^{k_e} \frac{k^2 dk}{\pi^2} \left( k^2 + m_e^2\right)^{1/2} \\
&= \frac{m_e^4}{8\pi^2} \left( (2t^2+1)^2 t(t^2 + 1)^{1/2} - \log(t + (t^2+1)^{1/2}) \right),
\end{align}
where $t = k_e / m_e$ with $k_e$ the electron Fermi momentum: 
\begin{equation}
    k_e = \frac{(3\pi^2 n_e)^{2/3}}{2m_e}.
\end{equation}
Note that the Fermi momentum is a function of the electron number density.

For a given baryon number density $n_B$, we have
\begin{equation}
 n_N = n_B / A, \;\;\;\; n_e = Z n_N = n_B Z / A.
\end{equation}
Thus at any $n_B$ we can find the $\{Z,A\}$ that minimize $E_{tot}$, using known binding energies and extrapolating from experimental data where necessary. Known binding energies are taken from the \texttt{Mathematica} v12.0.0.0 function \texttt{IsotopeData} \cite{reference.wolfram_2020_isotopedata}.
The extrapolation is performed using the semi-empirical mass formula (SEMF) for the binding energy of nuclei~\cite{Weizscker:1935zz,1968fup..book.....A}: 
\begin{multline}
b_N(A,Z) = a_V A - a_S A^{2/3} - a_C \frac{Z(Z-1)}{A^{1/3}} +\\
- a_A \frac{(A-2Z)^2}{A} -  \delta(A,Z),
\end{multline}
where
\begin{equation}
    \delta(A,Z) = \begin{cases}
    0 & \textrm{$A$ is odd}, \\
    a_P A^{-1/2} & \textrm{$A$ is even and $Z$ is even}, \\
    -a_P A^{-1/2} & \textrm{$A$ is even and $Z$ is odd}.
    \end{cases}
\end{equation}
We take values for the SEMF coefficients following the least-squares fit in \cite{1968fup..book.....A}:
\begin{eqnarray}
    a_V &=& 15.8\,\textrm{MeV},\\
    a_S &=& 18.3\,\textrm{MeV},\\
    a_C &=& 0.714\,\textrm{MeV}, \\
    a_A &=& 23.2\,\textrm{MeV},\\
    a_P &=& 12.0\,\textrm{MeV}.
\end{eqnarray}

The pressure increases continuously with increasing depth, which implies that each change in nuclear species is accompanied by a discontinuity in the density. For this reason, when calculating the EoS, it is more convenient to take the pressure $p$ as the independent variable. At a fixed pressure, the quantity instead to be minimised is actually the chemical potential $\mu_B$:
\begin{equation}
 \label{chempot}
 \mu_B = \frac{E_{tot} + p}{n_B}.
\end{equation}
The pressure is given by
\begin{equation}
 \label{pressure}
 p = n_B^2 \frac{\partial(E_{tot}/n_B)}{\partial n_B},
\end{equation}
from which we obtain
\begin{equation}
\label{pressure_2}
    p = p_e + \frac{1}{3} W_L n_N,
\end{equation}
where the electronic pressure $p_e$ is
\begin{equation}
  \label{e.pe}
    p_e = n_e^2 \frac{\partial(E_{e}/n_e)}{\partial n_e}.
\end{equation}
One can then show that the chemical potential is given by
\begin{equation}
\label{chempot_2}
 \mu_B = \left(W_N + \frac{4}{3}W_L + Z \mu_e \right) / A,
\end{equation}
where $\mu_e$ is the electron chemical potential, defined analogously to Eq.~\eqref{chempot}:
\begin{equation}
    \mu_e = \frac{E_e+p_e}{n_e}.
\end{equation}

At a given pressure, therefore, for an assumed $Z, A$ we can use Eq.~\eqref{pressure_2}, (\ref{e.pe}), (\ref{e.crustEe}) and (\ref{e.WL}) to solve for the electron density (which, like the pressure, is a continuous function of depth), remembering that $n_e = Z n_N$. Then for a particular electron density one can evaluate the chemical potential using Eq.~\eqref{chempot_2}, and then minimize over $\{ {Z, A} \}$ to find the equilibrium nuclear species. Once the species is known, the mass/energy density can be easily obtained and hence the relationship between pressure and energy density $p(\epsilon)$.

\subsection{Crust Model for a Mirror Neutron Star}

Extending these results to mirror sector matter amounts to finding the correct expression for $E_{tot}$ for larger values of the quark masses. The electron mass is scaled in the same ratio as the quark masses:
\begin{equation}
    \frac{m_e'}{m_e} = \frac{m_q'}{m_q} = \frac{f}{v},
\end{equation}
while the nucleon mass is scaled as for the core using the information in Table~\ref{tab:params}.

To obtain mirror binding energies and hence mirror nucleus masses, we construct a table of SM binding energies $b_N(A,Z)$ for the different values of $A,Z$ (using either experimental data or the semi-empirical mass formula, as explained above) and rescale those values by $\Lambda'_\mathrm{QCD}/\Lambda_\mathrm{QCD}$. 
We expect this naive dimensional rescaling to capture the lowest-order change in mirror binding energies compared to the SM, but of course the detailed binding energies must be treated as unknown. 
To evaluate the sensitivity of our results to this significant uncertainty, we repeat the derivation of neutron star structure and observables (explained in the next section) for many different crust models where for each value of $A, Z, f/v$, each individual binding energy in the table of $b_N(A,Z)$ is multiplied by a random factor in the range $(0.5, 2)$, in addition to the rescaling with $\Lambda'_\mathrm{QCD}$. We find that this affects the results of our analysis at the percent-level or less. Therefore, the precise details of mirror nuclear binding energies can be safely assumed to not dominantly change the structure of neutron stars at the level of precision relevant to our investigation.

\subsection{Interpolation Region}

Finally the intermediate region between the crust and core is covered by an interpolation function. Our procedure is to find an interpolation for both $p(n_B)$ and $\epsilon(n_B)$ between the point of the neutron drip and nuclear saturation density. First, we note that both the neutron drip and nuclear saturation density are expected to scale in some way with mirror quark mass. Following \cite{Baym:1971pw}, we can find the density/pressure of the neutron drip for each $f/v$ by solving the condition $\mu_B - m_N = 0$, where the neutron mass is scaled appropriately (see above). We find that the baryon number density $n_B$ at which neutron drip occurs scales approximately with $(\Lambda_{QCD}')^3$. Similarly, we choose to scale nuclear saturation density with $(\Lambda_{QCD}')^3$.

The interpolation is chosen such that the EoS is continuous, both at neutron drip and at nuclear saturation density. We also enforce that both $p$ and $\epsilon$ are strictly monotonically increasing with density. This is achieved using a Fritsch-Carlson monotonic cubic interpolation \cite{10.2307/2156610}. We note that in our core model, the gradient of $p(n_B)$ changes very rapidly as $n_B$ approaches saturation density. This means that the interpolating function below nuclear saturation can change dramatically depending on the precise density at which it is matched to the core model. In fact we find that the slope of the EoS at nuclear saturation can impact on the radius of intermediate mass neutron stars. Thus, rather than match at $n_B^{sat}$, we match at the slightly higher density $n_B = r_B n_B^{sat}$, with $0 < r_B - 1 \ll 1$, where the exact value of $r_B = 1.02$ is chosen such that the resulting mass-radius relationship agrees with observational constraints from NICER~\cite{Riley:2019yda,Miller:2019cac} and LIGO/VIRGO~\cite{TheLIGOScientific:2017qsa,Abbott:2018exr,Abbott:2018wiz}. 
When we extend the model to mirror QCD, where there are no observational constraints to fit to, we simply match at the same density ratio $(n_B / n_B^{sat}) = r_B = 1.02$ as in the SM case. Future work could explore how the properties of  liquid nuclear matter and nuclear pasta phases change for mirror matter (beyond the simple rescaling with $\Lambda_\mathrm{QCD}$ that we account for), which would further improve the robustness of our mirror EoS.

\begin{figure}
    \centering
    \includegraphics[clip=true,width=7cm]{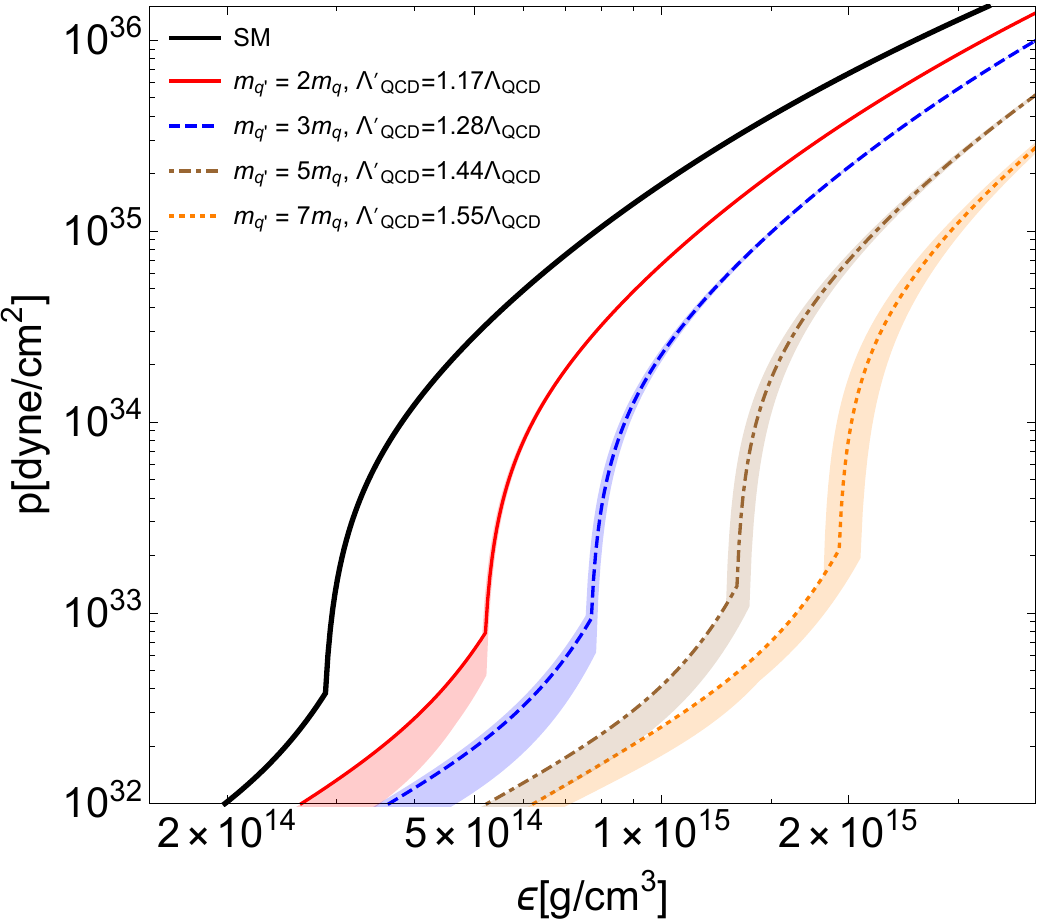}
    \caption{EoSs (pressure vs. energy density) of Mirror Neutron Stars compared to our SM Neutron Star. The kink marks the transition between  crust and core. 
    }
    \label{fig:EoS}
\end{figure}

\section{Modeling (Mirror) Neutron Stars in General Relativity}
Once an EoS is known, the structure of a neutron star (mirror or not) and its gravitational field is determined by the Einstein equations: 10 coupled, non-linear, partial differential equations \cite{1967ApJ...150.1005H}. 
Since all observed neutron stars are rotating slowly relative to the Keplerian mass-shedding limit, one can expand the Einstein equations in a slow-rotation approximation. Doing so, assuming an isolated star, the Einstein equations reduce to the Tolman-Oppenheimer-Volkoff (TOV) equations at zeroth-order in the slow-rotation approximation. The solution to these equations yields the mass and radius of the star, for a given central density; choosing a set of central densities, one can construct mass-radius curves. At first order in rotation, the solution of a single partial differential equation yields the moment of inertia of the star. At second order in rotation, the solution of a set of partial differential equations yields the (rotational) quadrupole moment of the star. 

When a neutron star is not isolated, but rather is in the presence of a companion, the solution to the Einstein equations reveal how the neutron star deforms~\cite{Thorne:1997kt}. This deformation is encapsulated in the ($\ell=2$, tidal) Love number, 
which measures the induced quadrupole moment due to a quadrupolar tidal force.
The Love number is calculated by perturbing the Einstein equations due to an external tidal field, and solving these equations to first order in the tidal perturbation. The Love number depends on the mass and radius 
of the unperturbed star, since stars with a lower compactness (ratio of mass to radius) are easier to deform, leading to a larger Love number. If the companion to the neutron star is also a neutron star, then the secondary will  possess its own Love number.     

The quantities mentioned above (mass, radius, moment of inertia, quadrupole moment and Love number) are astrophysical observables.
For example, gravitational wave observations of binary neutron stars are sensitive to the masses of the stars and their Love numbers, because these quantities affect the binding energy of the binary and  the rate  it inspirals~\cite{Flanagan:2007ix}. 
Therefore, mirror neutron star properties could be measured in gravitational wave observations of mirror neutron star binary mergers. 
Similarly, radio observations of binary pulsars are sensitive to the masses of the binary and their moment of inertia, because the latter generates precession of the orbital plane (through spin-orbit coupling interactions), which imprints on the arrival times  of the pulses~\cite{Lattimer:2004nj}.  
Therefore, measuring mirror neutron star properties might be possible via radio observations of asymmetric binaries, consisting of a SM pulsar and a mirror neutron star. 

The measurement of these observables is complicated by parameter degeneracies that exist in the models used to fit the data. For example, the Love numbers of two neutron stars in a binary affect the gravitational wave model in a similar manner (i.e.~they both first enter at 5th order in the post-Newtonian expansion used to create these models~\cite{Flanagan:2007ix}).
One can measure a certain combination of them, but not necessarily their individual values. Certain approximately universal relations have been discovered that help break these degeneracies. For example, the I-Love-Q relations between moment of inertia, Love number, and rotational quadrupole moment 
are insensitive at the $< 1\%$ level to variations in the EoS~\cite{Yagi:2013bca,Yagi:2013awa}. Another example are binary Love relations between two Love numbers in a binary system have been shown to be insensitive to variations in the EoS to better than 10\%~\cite{Yagi:2015pkc,Yagi:2016bkt}. 
These approximately universal relations are important because they can break degeneracies in parameter estimation \cite{Yagi:2015pkc,Yagi:2016bkt,Chatziioannou:2018vzf,Abbott:2018exr}  and are model independent and EoS-insensitive tests of general relativity~\cite{Yagi:2013bca,Yagi:2013awa,Yagi:2016bkt,Silva:2020acr}.

%
We here provide a pedagogical review of this calculation, aimed at non-gravitational physicists, which is identical to the corresponding calculation for SM neutron stars, following~\cite{1967ApJ...150.1005H,1968ApJ...153..807H,Yagi:2016bkt}. 
Henceforth, we use the conventions $G = 1 = c$ and metric signature $\left(-,+,+,+\right)$.

\subsection{Spacetime Metric of a (Mirror or SM) Slowly-Rotating Neutron Star}
\label{sec:expand_metric}

We consider an isolated neutron star, whose interior is described by some EoS (made of mirror matter or SM matter), and that rotates uniformly with angular velocity $\Omega$. 
We assume the star rotates slowly, $\Omega^2 R_* \ll GM_*/R_*^2$, where $R_*$ and $M_*$ are the star's radius and mass, such that we can expand all expressions perturbatively to $\mathcal{O}\left(\Omega^2\right)$. 
The spacetime metric to this order in slow rotation takes the form~\cite{1967ApJ...150.1005H,1968ApJ...153..807H,Yagi:2013awa}
\begin{equation}\label{eq:metric}
\begin{split}
	ds^2 = & -e^{\nu}\left(1+2h\right)dt^2
	+e^{\lambda}\left(1+\frac{2m}{r-2M}\right)dr^2 \\
	& +r^2\left(1+2k\right)\left(d\theta^2+\sin^2\theta\left(d\phi-\left(\Omega-\omega\right) dt\right)^2\right),
\end{split}
\end{equation}
where $\nu(r)$ and $\lambda(r)$ are metric functions of $\mathcal{O}\left(\Omega^0\right)$, $\omega(r,\theta)$ is of $ \mathcal{O}\left(\Omega^1\right)$, and $(h(r,\theta), m(r,\theta), k(r,\theta))$ are of $\mathcal{O}\left(\Omega^2\right)$. The quantity $M$ here is a function of radius, and thus not to be confused with the total mass of the star, which we shall call the \textit{enclosed mass} and will define as
\begin{equation}\label{eq:}
	M\left(r\right) = \frac{r}{2} \left(1-e^{-\lambda} \right)\,.
\end{equation}
The functions $M$, $\nu$ and $\lambda$ are independent of polar angle because, to ${\cal{O}}(\Omega^0)$ the spacetime is spherically symmetric.

The neutron star described by the metric above will deform due to rotation, becoming an ellipsoid instead of a sphere when $\Omega \neq 0$. For this reason, we cannot define the surface of the rotating object via a condition like $r = R_*$, and instead, the surface is a function of the polar angle $\theta$. This, in turn, implies that the energy density at the (spherical) surface of the non-rotating star will not vanish everywhere on the (ellipsoidal) surface of the rotating star. But in order for the perturbative scheme to work, we must be able to expand the energy density as $\epsilon = \epsilon_0 +  \epsilon_1 + \epsilon_2 + {\cal{O}}(\Omega^3)$, with $\epsilon_n = {\cal{O}}(\Omega^n)$ and thus $\epsilon_{n+1} \ll \epsilon_n$, which is simply not possible if $\epsilon_0 = 0$. In order to avoid this complication, we can perform the coordinate transformation
\begin{equation}\label{eq:rtoR}
\begin{split}
	& r(R,\Theta) = R + \xi \left( R, \, \Theta\right)\,, \qquad
	 \theta = \Theta\,,
\end{split}
\end{equation}
defined such that $\epsilon[r(R,\Theta),\theta(\Theta)] = \epsilon(R) = \epsilon_0(R)$, i.e.~the coordinate $R$ is the radius at which the density of a given isodensity contour of the rotating configuration equals the density of an isodensity contour in the non-rotating configuration. With this transformation, the condition $r=r(R_{*,0},\Theta)$ defines the surface of the star, with $R_{*,0}$ the stellar radius of the non-rotating configuration and the function $\xi(R,\Theta)$ determining the oblateness of the rotating star, which must be of $\mathcal{O}\left(\Omega^2\right)$ (since oblateness is invariant under time-reversal). Heuristically, one can think of $R$ as the radial coordinate of a non-rotating configuration, $r$ as the radial coordinate of the rotating configuration, and $R_*$ as the radius of the rotating star as measured in the $R$ coordinate system. 

As discussed above, we will expand all quantities in a slow-rotation expansion, and thus, it is necessary to establish some notation. In general, we will first transform all quantities from $r \to R$ and perform a Taylor expansion, so that $g(r,\theta) = g(R+\xi,\theta) = g(R,\theta) + \xi \partial_r g(R,\theta)$. Then, we will expand all remaining quantities in a slow-rotation expansion such as $g(R,\Theta) = \sum_n g_n(R,\Theta)$, where $g_n = {\cal{O}}(\Omega^n)$. The resulting expressions are of the form $g(r,\theta) = g_0(R,\theta) + g_1(R,\theta) + g_2(R,\theta) + \xi \partial_r g_0(R,\theta)$, and there are clearly two terms of ${\cal{O}}(\Omega^2)$ because $\xi = {\cal{O}}(\Omega^2)$. Moreover, we will find it convenient to expand the angular dependence in Legendre polynomials of cosine polar angle: $g_n(R,\theta) = \sum_{\ell} g_{n \ell}(R) P_{\ell}(\cos\theta)$. Because of the boundary conditions at $r \to 0$ and $r \to \infty$, and the invariance under $\Omega \to -\Omega$, the metric coefficient $\omega$ is non-zero only when $l=1$ and $h, m$, and $k$ are non-zero only when $l=0$ or $2$ (see for further details~\cite{1967ApJ...150.1005H,1968ApJ...153..807H}).

\subsection{Einstein Equation to Zeroth Order in Slow Rotation}
\label{sec:0th_Eeq}

To zeroth order in rotation, the Einstein equations, $G_{\mu\nu} = 8 \pi T_{\mu\nu}$, 
reduce to
\begin{subequations}
\begin{align}
&\frac{d M}{dR} = 4 \pi \epsilon R^2\\
&\frac{d p}{d R} = -\left(\epsilon+p\right)\frac{M+4\pi p R^3}{R\left(R-2M\right)} \\
&\frac{d \nu}{d R} = 2 \frac{M+4\pi p R^3}{R\left(R-2M\right)}
\end{align}\label{eq:TOV} 
\end{subequations}

where recall that $M$ is a function of radius that is related to the $(r,r)$ component of the metric, while $\nu$ is another function of radius related to the $(t,t)$ component of the metric [see Eq.~\eqref{eq:metric}].
We have here omitted the subscript 0 (which denotes the rotation order) for simplicity. There is also no need here for a Legendre expansion, since at this order the spacetime is spherically symmetric. The second equation in the set above is known as the Tolman-Oppenheimer-Volkoff (TOV) equation. The above three equations, together with an EoS $p=p(\epsilon)$ defines a closed set of ordinary differential equations. The EoS is the only ingredient in this set that determines whether one is considering SM neutron stars or mirror neutron stars. 

This set of equations must now be solved both in the interior of the star and in the exterior (where $p = 0 = \epsilon$), ensuring continuity at the surface, which (to zeroth-order in rotation) we will say is located at $R=R_{*,0}$. 
The exterior equations can be solved exactly to find $p_{\rm ext} = 0 = \epsilon_{\rm ext}$, $\nu_{\rm ext} = \ln(1 - 2 M_{*,0}/R)$ and $M_{\rm ext} = M_{*,0}$, where $M_{*,0}$ is a constant of integration (to be identified later with the total mass of the star to zeroth-order in rotation). The interior solution must be obtained numerically because the EoS is usually provided as a numerical table. To do so, one first needs to find boundary conditions at the center of the star $R_c$, which can be obtained by solving Eqs.~\eqref{eq:TOV} asymptotically about $R_c \ll R_{*,0}$. Doing so, one finds
\begin{subequations}\label{eq:BC_TOV}
	\begin{align}
	& M\left(R_c\right) = \frac{4 \pi}{3} \epsilon_c R_c^3 + \mathcal{O}\left(R_c^5\right)\\
	& p\left(R_c\right) = p_c - 2 \pi \left(\epsilon_c+p_c\right) \left(\epsilon_c/3+p_c\right) R_c^2+\mathcal{O}\left(R_c^4\right)\\
	& \nu\left(R_c\right) = \nu_c + 4 \pi \left(\epsilon_c/3+p_c\right) R_c^2+\mathcal{O}\left(R_c^5\right)\,,
	\end{align}
\end{subequations}
where $\epsilon_c$ and $\nu_c$ are two more constants of integration, while $p_c = p(\epsilon_c)$. Given a choice of the central density $\epsilon_c$, one can then use the above boundary conditions to integrate $M_{\rm int}$ and $p_{\rm int}$ from the central radius $R_c$ (a small nonzero but otherwise arbitrary starting point for the numerical integration) to the stellar surface $R_{*,0}$, with the latter defined via $p_{\rm int}(R_{*,0}) = \varepsilon \, p_c$, with $\varepsilon$ also arbitrary as long as $\varepsilon \ll 1$.
The constant of integration $\nu_c$ is then determined by requiring continuity at the surface, $\nu_c = \ln(1 - 2 M_{*,0}/R_{*,0})$, where $M_{*,0} = M(R_{*,0})$.

We have carried out these calculations for the mirror neutron star EOS described earlier, which leads to the mass-radius curves shown in Fig.~\ref{fig:MRplotfront}. 
Each choice of central density then yields a \textit{single} neutron star, with a given mass and a given radius (both determined from the integration of the equations as described above). Choosing a sequence of values for the central density then yields a mass-radius curve. 
For ease of reading, in that figure we use the symbols $M$ and $R$ to refer to the mass and radius of the star at zeroth-order in rotation, which we referred to in this subsection as $M_{*,0}$ and $R_{*,0}$ (none of which ought to be confused with the radial coordinate $R$ used earlier). All integrations are done with $R_c = 10$ cm and $\varepsilon = 10^{-8}$, although we have checked that our results do not change appreciably (i.e.~by more than $0.1$ km) if we vary these integration parameters.

\subsection{Einstein Equation to First Order in Slow Rotation}
\label{sec:1st_Eeq}

The only relevant equation at first order in rotation comes from the $\left(t,\phi\right)$ component of the Einstein equations. Since this component transforms as a vector in the slow-rotation expansion, we know that $\omega_1(r,\theta)$ can be decomposed in terms of vector spherical harmonics, i.e.~we know that $\omega_1(r,\theta) = \sum_{\ell} \omega_{1,\ell} P_{\ell}'(\cos{\theta})$. Moreover, we know that $\omega$ must be odd under time reversal (because it must be ${\cal{O}}(\Omega)$), and this then implies that only the $\ell=1$ mode is excited. The perturbed Einstein equations then yield an equation for $\omega_{1,1}$, namely
\begin{equation}\label{eq:1st_order}
	\frac{d^2 \omega_1}{d R^2} +\frac{4}{R}\left(1-\frac{\pi R^2\left(\epsilon+p\right)}{1-2M/R}\right)\frac{d \omega_1}{d R}- \frac{16\pi \left(\epsilon+p\right)}{1-2M/R}\omega_1= 0\,,
\end{equation}
where we have just dropped the rotation subindex (since this entire subsection is about ${\cal{O}}(\Omega)$ perturbation), but kept the $\ell$-harmonic index.
The energy density and pressure here are zeroth-order in rotation because again $\omega_1 = {\cal{O}}(\Omega)$.

As in the ${\cal{O}}(\Omega^0)$ case, we must now solve the above equation in the interior and in the exterior of the star, ensuring continuity and differentiability at the surface. The exterior solution is simply
\begin{equation}
  \omega_1^{ext} = \Omega-{2 S}/{R^3} 
\end{equation}
where $S$ is a constant of integration (later to be identified with the magnitude of the spin angular momentum), and the other integration constant is set to zero by requiring asymptotic flatness. In the interior, we must once more solve the above differential equation numerically, with a boundary condition obtained by solving this equation asymptotically in $R_c \ll R_{*,0}$. Doing so, one finds 
\begin{equation}
	\omega_1\left(R_c\right) = \omega_c\left[1+\frac{8\pi}{5}\left(\epsilon_c+p_c\right)R_c^2\right]\,,
\end{equation}
where $\omega_c$ is another integration constant and we have set the second integration constant to zero by requiring smoothness at the center. Given a rotation frequency $\Omega$, one can then find the values of $\omega_c$ and $S$ that lead to a continuous and differentiable solution at $R_{*,0}$, for example through a numerical shooting method or by exploiting the homogeneity and scale invariance of Eq.~\eqref{eq:1st_order} (see e.g.~\cite{Yagi:2013awa} for more details). Once the correct $(\omega_c,S)$ has been found for a choice of $\Omega$, one can read out the moment of inertia via $I := S/\Omega$. Note that the moment of inertia is actually independent of $\Omega$, since Eq.~\eqref{eq:1st_order} is scale invariant, and thus, we can recast it as an equation for $
\bar{\omega}_1:=\omega_1/\Omega$, so the shooting problem becomes one of determining $\bar{\omega}_c$ and $I$ directly. 

We have carried out these calculations for the mirror neutron star EOS described earlier, which leads to the moment-of-inertia-mass curves shown in Fig.~\ref{fig:mirrorobservablesplot} (top panel). The parameter along each of these lines is the central density, since as we explained above, the moment of inertia is independent of $\Omega$. Figure~\ref{fig:mirrorobservablesplot} actually plots the \textit{dimensionless} moment of inertia (i.e.~$I/M_{*,0}^3$), since $I$ has units of $[\rm{Length}]^3$. All integrations are done with the same choices of $R_c$ and $\varepsilon$ as in the ${\cal{O}}(\Omega^0)$ calculation.

\subsection{Einstein Equation to Second Order in Slow Rotation}
\label{sec:2nd_Eeq}

At second order in $\Omega$, one finds differential equations for the metric functions $m$, $k$ and $h$, as well as for the oblateness parameter $\xi$. Since these quantities behave as scalars under rotations, they can be expanded in scalar spherical harmonics, and moreover, since they must be even under rotation, only the $\ell=0$ and $\ell=2$ modes are excited. The $\ell=0$ mode will lead to a $\Omega^2$ correction to the radius and the mass of the star, which we do not consider in this paper. The $\ell=2$ mode will allow us to calculate the rotation-induced quadrupole deformation of the star. Since $(m,k,h,\xi)$ are all of ${\cal{O}}(\Omega^2)$, we will suppress the rotation sub-index, but we will keep the $\ell$-harmonic index. 

The Einstein equations can be simplified through the stress-energy conservation equation. The latter allows us to find an expression for $\xi_2$ in terms of ${\cal{O}}(\Omega^0)$ quantities ($\nu, \lambda, M, p, \epsilon$) and ${\cal{O}}(\Omega)$ quantities ($\omega_1$). Using this expression in the Einstein equations, one then finds the following coupled set of ordinary differential equations for $(
h_2,k_2)$:  
\begin{align}
\begin{split}
\frac{d h_2}{d R} = 
&-\frac{R-M+4\pi pR^3}{R} e^\lambda \frac{d k_2}{d R}+\frac{3-4 \pi \left(\epsilon+p\right)R^2}{R} e^\lambda h_2 \\
&+ \frac{2}{R}e^\lambda k_2+\frac{1+8\pi p R^2}{R^2}e^{2\lambda}m_2+\frac{R^3}{12}e^{-\nu}\left(\frac{d \omega_1}{d R} \right)^2 \\
&-\frac{4 \pi \left(\epsilon+p\right)R^4\omega_1^2}{3R}e^{-\nu+\lambda}\,,
\end{split}\label{eq:2orderQeq}
\end{align}

\begin{align}
\begin{split}
\frac{d k_2}{d R} = & -\frac{d h_2}{d R}+ \frac{R-3M-4\pi p R^3}{R^2}e^\lambda h_2 \\ &+\frac{R-M+4\pi p R^3}{R^3} e^{2\lambda}m_2\,, \label{eq:dk_2}   
\end{split}
\end{align}
with the constraint
\begin{equation}
\begin{split}
    m_2  =
    & -R e^{-\lambda}h_2 \\
    &+\frac{1}{6}R^4 e^{-\nu-\lambda}\left(R e^{-\lambda}\left(\frac{d \omega_1}{d R} \right)^2+16 \pi\left(\epsilon+p\right)R \omega_1^2\right)\,. \label{eq:m_2}
\end{split}
\end{equation}

\vspace{-0.2cm}

As before, these equations must be solved in the interior and in the exterior of the star, ensuring continuity at the surface. In the exterior, one finds
\begin{widetext}
\begin{subequations}\label{eq:2orderSolext}
\begin{align}
	h_2^{ext} & = \frac{S^2}{M_{\ast,0} R^3}\left(1+\frac{M_{\ast,0}}{R}\right)-c_1\frac{3 R^2}{M_{\ast,0}\left(R-2M_{\ast,0}\right)}\left(1-3\frac{M_{\ast,0}}{R}+\frac{4M_{\ast,0}^2}{3R^2}+\frac{2M_{\ast,0}^3}{3R^3}+\frac{R}{2M_{\ast,0}}f_{schw}^2\ln f_{schw}\right) \\
	k_2^{ext} & = \frac{-S^2}{M_{\ast,0} R^3}\left(1+\frac{2M_{\ast,0}}{R}\right)+c_1\frac{3R}{M_{\ast,0}}\left(1+\frac{M_{\ast,0}}{R}-\frac{2M_{\ast,0}^2}{3R^2}+\frac{R}{2M_{\ast,0}}\left(1-\frac{2M_{\ast,0}^2}{R^2}\right)\ln f_{schw}\right) \\
	m_2^{ext} & = \frac{-S^2}{M_{\ast,0} R^3}\left(1-7\frac{M_{\ast,0}}{R}+10\frac{M_{\ast,0}^2}{R^2}\right)+c_1\frac{3R^2}{M_{\ast,0}}\left(1-\frac{3M_{\ast,0}}{R}+\frac{4M_{\ast,0}^2}{3R^2}+\frac{2M_{\ast,0}^3}{3R^3}+\frac{R}{2M_{\ast,0}}f_{schw}^2\ln f_{schw}\right)
\end{align}
\end{subequations}
\end{widetext}
where 
\begin{equation}
    f_{schw} = 1- \frac{2M_{\ast,0}}{R}, 
\end{equation}
$c_1$ is an integration constant and we have set the other integration constant to zero by requiring asymptotic flatness. The quadrupole moment can be read off from the far-field expansion of the $(t,t)$ component of the metric, $g_{tt} \sim -1 + 2 M_{*}/R - 2 Q/R^3 P_2(\cos{\theta}) + {\cal{O}}(R^{-4})$, and since $h_2$ determines $g_{tt}$ at this order, we then find that 
\begin{equation}
    Q = -\left(\frac{S^2}{M_\ast}+\frac{8}{5} c_1 M_\ast^3\right).
\end{equation}
In the interior, one must solve the differential equations numerically, starting with a boundary condition at the center, which is obtained asymptotically in $R \ll R_{*,0}$: $h_2(R_c) = B \cdot R_c^2 + {\cal{O}}(R_c^4)$ and $k_2(R_c) = -B \cdot R_c^2 + {\cal{O}}(R_c^4)$, where $B$ is an integration constant. Given a choice of central density $\epsilon_c$ and angular frequency $\Omega$, one must then find the constants $c_1$ and $B$ that will yield a solution for $k_2$ and $h_2$ that is continuous at the surface $R_{*,0}$. Once more, this can be achieved through a numerical shooting algorithm, or by exploiting the properties of the differential system.

We have carried out these calculations for the mirror neutron star EoS described earlier, which leads to the quadrupole-momment-mass curves shown in Fig.~\ref{fig:mirrorobservablesplot} (middle panel). The parameter along each of these lines is again the central density, since we plot here the dimensionless quadrupole moment $Q M/S^2$, which is independent of $\Omega$. All integrations are done with the same choices of $R_c$ and $\varepsilon$ as in the ${\cal{O}}(\Omega^0)$ calculation.

\subsection{Einstein Equations for a Tidally Perturbed Star}
\label{sec:Love_Eeq}

We are now interested in understanding the structure of a neutron star that is tidally modified by some external gravitational perturbation. For this calculation, we do not care about rotation at all, so the background spacetime is spherically symmetric. Tidal deformations can of course be excited on rotating stars, but those deformations will not affect what we calculate below to leading-order in a slow-rotation expansion. The metric ansatz, however, happens to be identical to that of the slow-rotation approximation [Eq.~\eqref{eq:metric}], except that $\omega_1 = 0 = \Omega$, $(m,k,h)$ are first order in the tidal deformation, and $\xi$ is induced by the tidal deformation instead of rotation; it is also conventional to expand the $(m,k,h)$ metric functions in spherical harmonics instead of Legendre polyonomials, but this only introduces a constant factor. Because of this, the Einstein equations at zeroth-order in tidal deformation are identical to those obtained at zeroth-order in slow-rotation [Eqs.~\eqref{eq:TOV}], while the Einstein equations at first-order in tidal deformation are identical to those obtained at second-order in slow-rotation [Eqs.~(\ref{eq:2orderQeq}-\ref{eq:dk_2})  with $\omega_1 = 0 = \Omega$]. In practice, this implies $m_2 = -R e^{-\lambda} h_2$, which allows us to decouple the equations and find a second order differential equation for $h_2$:
\begin{widetext}
\begin{equation}
\label{eq:h2tid}
\frac{d^2 h_2}{dR^2} + \left[\frac{2}{R}+\left(\frac{2M}{R^2}+4\pi R\left(p-\epsilon\right)e^\lambda\right)\right]\frac{d h_2}{d R}  
-\left[ 
\frac{6 e^\lambda}{R^2}
-4\pi\left(5\epsilon+9p+\left(\epsilon+p\right)\frac{d\epsilon}{dp}\right)e^\lambda+\left(\frac{d\nu}{dR}\right)^2
\right]h_2 = 0\,.
\end{equation}
\end{widetext}

As before, this equation must be solved in the exterior and in the interior of the star, requiring continuity and differentiability at the surface $R_{*,0}$. In the exterior, one finds the solution~\cite{Hinderer:2007mb}:
\begin{equation}\label{eq:tidalsolext}
\begin{split}
	&h_2^{ext} = C_2\left(\frac{R}{M_{\ast,0}}\right)^2f_{schw} 
	 -C_1\frac{R^2}{M_{\ast,0}^2}f_{schw} \\
	 & \left[\frac{2M_{\ast,0}\left(R-M_{\ast,0}\right)\left(3R^2-6M_{\ast,0} R-2M_{\ast,0}^2\right)}{R^2\left(R-2M_{\ast,0}\right)^2} 
	+3 \ln f_{schw} \right]\,, \\
\end{split}
\end{equation}
where $C_1$ and $C_2$ are integration constants and recall that we previous defined $f_{schw} = 1- {2M_{\ast,0}}/{R}$. 
Unlike in the slow-rotation case, we are here interested in a tidally perturbed spacetime, which is only formally defined far away from the source of the perturbation $R \ll R_{\rm source}$. This is why we have not imposed asymptotic flatness to eliminate one of the integration constants. The interior solution must be found numerically again, with boundary conditions obtained by solving Eq.~\eqref{eq:h2tid} asymptotically about $R \ll R_{*,0}$; these conditions happen to be the same as those found in the slow-rotation case, namely $h_2(R_c) = B R_c^2 + {\cal{O}}(R_c^4)$. One must then find the constants of integration $B$ and $C_1/C_2$ such that the spacetime is continuous and differentiable at the surface. Note that because of the scale-invariance of Eq.~\eqref{eq:h2tid}, only the ratio of $C_1$ and $C_2$ (and the integration constant of the interior solution $B$) is necessary to find a smooth solution.

With the solution at hand, one then wishes to extract observable quantities, like the tidal Love number. A multipolar expansion of the metric in the buffer zone, defined as the spatial region $R_{*,0} \ll R \ll R_{\rm ext}$ (with $R_{\rm ext}$ the radius of curvature of the source of the external field), yields~\cite{Hinderer:2007mb}:
\begin{equation}\label{eq:tidal_metric}
\begin{split}
	-\frac{1+g_{tt}}{2}  = & -\frac{M_{\ast,0}}{R}- \frac{3 Q_{ij}^{(tid)}}{2 R^3} \left(n^i n^j - \frac{1}{3} \delta^{ij} \right) + {\cal{O}}\left(\frac{1}{R^4}\right) \\ 
	&+ \frac{1}{2} {\cal{E}}_{ij} R^2 n^i n^j + {\cal{O}}\left(R^3\right)\,,
\end{split}
\end{equation}
where ${\cal{E}}_{ij}$ is the (quadrupole) tidal tensor field, 
$Q^{(tid)}_{ij}$ is the corresponding tidally-induced and traceless quadrupole moment tensor, $n^i = x^i/R$ is a field-point unit vector. We can rewrite these tensors in a spherical harmonic decomposition ${\cal{E}}_{ij} = \sum_m {\cal{E}}_m {\cal{Y}}_{ij}^{2m}$, with $Y_{2m} = {\cal{Y}}_{ij}^{2m} n^i n^j$ and similarly for $Q_{ij}^{(tid)}$, and then define the (quadrupole) 
tidal deformability $\lambda := -Q_m/{\cal{E}}_m$. Comparing this to the buffer zone expansion of the metric ansatz of Eq.~\eqref{eq:metric}
\begin{equation}
\begin{split}
    &-\frac{1+g_{tt}^{ext}}{2} =  -\frac{M_{\ast,0}}{R} + h_2^{ext} Y_{2m}(\theta) \\
    &=  -\frac{M_{\ast,0}}{R} +\left[ \frac{16}{5} C_1 \frac{M_{\ast,0}^3}{R^3} + \mathcal{O}\left(\frac{1}{R^4}\right) + C_2\frac{R^2}{M_{\ast,0}^2} + \mathcal{O}\left(R^3\right) \right]\times\\
   &\times Y_{2m}(\theta)
\end{split}
\end{equation}
means that the tidal deformability $\lambda$ is fully determined by the ratio $C_1/C_2$. In fact, one can write down the tidal deformability entirely in terms of the exterior solution $h_2^{ext}$ and its derivative evaluated at the surface of the star, namely
\begin{equation}
\begin{split}\label{eq:tidalLove}
	\Lambda &:= \frac{\lambda}{M_{*,0}^5} = \frac{16}{15}\frac{C_1}{C_2} \\
	 &=\frac{16}{15} \left\{\left(1-2 {C}\right)^2\left[2-y+2 {C}\left(y-1\right)\right]\right\} \left\{
	 2 {C}\left[6-3y+3 {C}(5y-8)\right] \right.
	  \\
	 &+ 4 {C}^3 \left[ 13-11y+  {C}(3y-2) + 2 {C}^2(1+y) \right] \\ 
	 & \left. +3(1-2 {C})^2\left[2-y+2 {C}(y-1)\right] \ln\left(1-2 {C}\right) 
	 \right\}^{-1}\,,
\end{split}    
\end{equation}

where the compactness defined via ${C} = M_{\ast,0}/R_{\ast,0}$ is the compactness and $y = (R_{\ast,0}/h_2) \, ({d h_2}/{dR}) $ all evaluated at $R=R_{\ast,0}$.
    
We have carried out these calculations for the mirror neutron star EoS described earlier, which leads to the tidal-deformability-mass curves shown in Fig.~\ref{fig:mirrorobservablesplot} (bottom panel). The parameter along each of these lines is again the central density, and we here plot the dimensionless tidal deformability $\Lambda = \lambda/M_{*,0}^5$. Note that we have colloquially referred to $\Lambda$ as the Love number, while in reality it is the dimensionless tidal deformability (related to the Love number $k_2$ via $k_2 = (3/2) (\lambda/R_{*,0}^5) = (3/2) \Lambda (M_{*,0}/R_{*,0})^5$).  All integrations are done with the same choices of $R_c$ and $\varepsilon$ as in the ${\cal{O}}(\Omega^0)$ calculation.

\section{Results}
\subsection{Properties of Mirror Neutron stars}
\begin{figure}[htb]
    \includegraphics[clip=true,width=8cm]{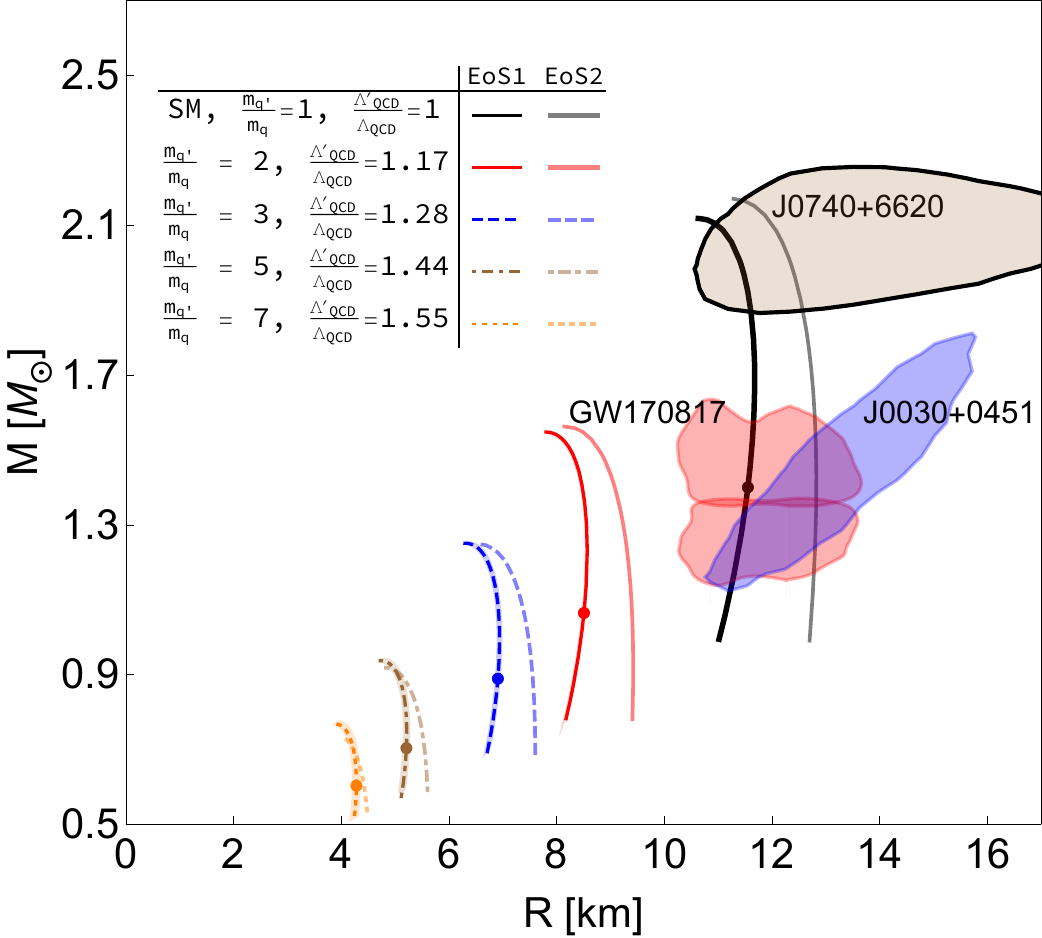}
    \caption{Mass-radius relations for mirror neutron stars with quark masses scaled  by $m_{q'}/m_q = f/v$. 
    The blue and red shaded regions correspond to $2\sigma$  confidence regions using X-ray observations of the neutron star J0030+0451 and gravitational wave observations of the neutron star merger GW170817, respectively~\cite{Abbott:2018exr,Riley:2019yda,Miller:2019cac}. The gray shaded region corresponds to the $1\sigma$ measurement of the mass of pulsar J0740+6620~\cite{Cromartie:2019kug}. 
    Solid lines (EoS1) represent results for the EoS from Sec.~\ref{sec:EoS}. 
     The shaded regions around these curves represent uncertainties in the $m_\pi$ dependence of various EoS input parameters (from chiral perturbation theory and Lattice QCD). Faint lines (EoS2) correspond to the modified EoS from Sec.~\ref{sec:altEoS}. The distance between faint (EoS2) and solid (EoS1) curves is meant to roughly represent systematic uncertainties. 
    }
    \label{fig:MRplotfront}
\end{figure}

We find  generically that the mass and radius of mirror neutron stars are smaller than their SM cousins, in Fig.~\ref{fig:MRplotfront}. This shift to lower masses and radii occurs because increased quark masses (and confinement scales) scale the pressure and energy density with $\left(\Lambda^\prime_{QCD}\right)^4$, which can also be understood from simple scaling arguments~\cite{Reisenegger:2015crq}. The minimum neutron star mass (the Chandrasekhar limit),  the maximum mass, and the  radius all scale with the nucleon mass as $\sim 1/m_n^2$. Remarkably, mirror neutron star mass-radius curves can be closely reproduced by rescaling both mass and radius in the SM mass-radius curve by $(m_{n'}/m_n)^{-1.9} \approx (\Lambda'_\mathrm{QCD}/\Lambda_\mathrm{QCD})^{-2.38}$. This close agreement to the naive scaling expectation implies that our results are robust to details of the hidden sector and depend mostly on the mirror confinement scale \cite{Narain:2006kx}.  

We conclude that mirror neutron stars are (electromagnetically) dark objects that have masses between $(0.5,1.3) M_\odot$ and radii between $(4,7) \, {\rm{km}}$ for $f/v \sim 3 - 7$. 
The smaller maximum mass of mirror neutron stars is not in conflict with binary pulsar observations, because mirror neutron stars are invisible to electromagnetic observations. This mass range is potentially the same as primordial black holes, except that mirror neutron stars have a non-zero Love number and a larger radius. Mirror neutron stars can be distinguished from other black-hole-like compact objects~\cite{Cardoso:2019rvt}, since their compactness never exceeds 
$C = GM/(c^2 R) \lesssim 0.3$, so their surface is far from their would-be horizons. 

\begin{figure}[htb]
    \includegraphics[clip=true,width=8cm]{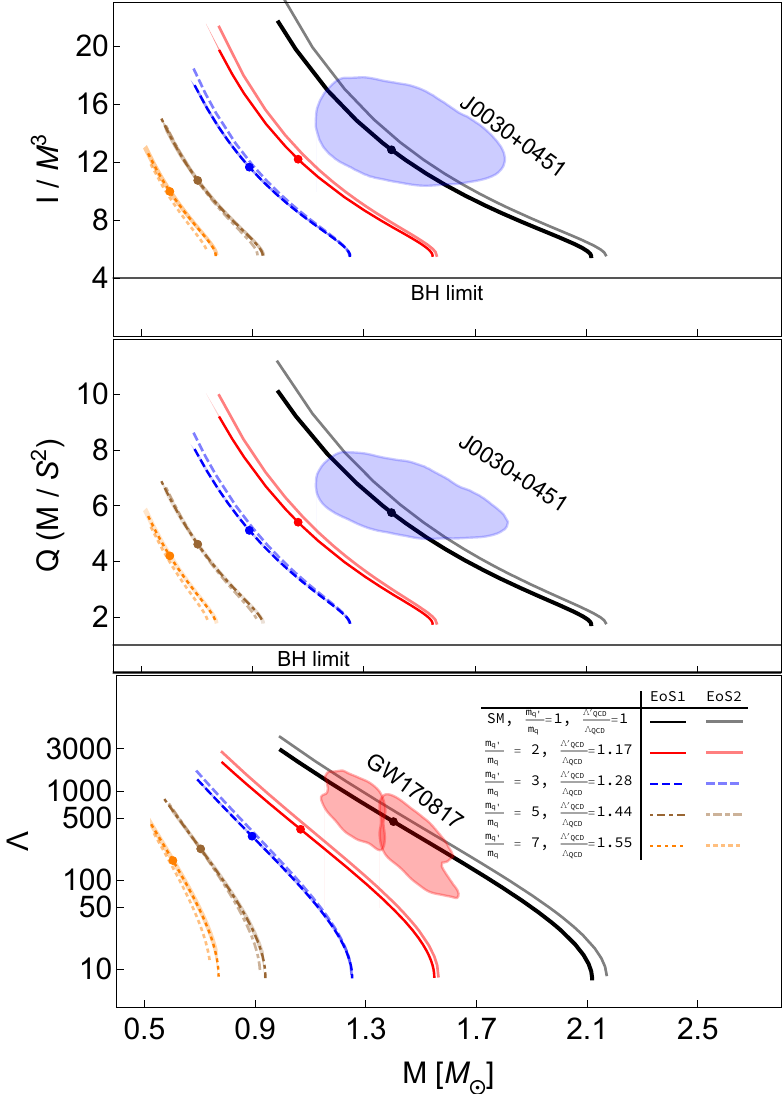}
    \caption{Moment of inertia $I$, quadrupole moment $Q$, and Love number $\Lambda$ of mirror neutron stars as a function of their mass. 
    The solid horizontal lines correspond to the black hole limit ($\Lambda=0$).  
    The two blue regions are obtained from the NICER observation and the approximate  compactness-I or compactness-Q universal relation. 
    Solid lines (EoS1) represent results for the EoS from Sec.~\ref{sec:EoS}. 
     The shaded regions around these curves represent uncertainties in the $m_\pi$ dependence of various EoS input parameters (from chiral perturbation theory and Lattice QCD). Faint lines (EoS2) correspond to the modified EoS from Sec.~\ref{sec:altEoS}. The distance between faint (EoS2) and solid (EoS1) curves is meant to roughly represent systematic uncertainties. }
    \label{fig:mirrorobservablesplot}
\end{figure}

Other observable properties of mirror neutron stars also differ from SM neutron stars. Figure~\ref{fig:mirrorobservablesplot} shows the moment of inertia, the (rotational) quadrupole moment, and the Love number of mirror neutron stars as functions of their mass. All  quantities shift to lower values for mirror neutron stars with higher quark masses. 

\subsection{Universal Relations}

\begin{figure*}
    \centering
    \includegraphics[clip=true,width=5.5cm]{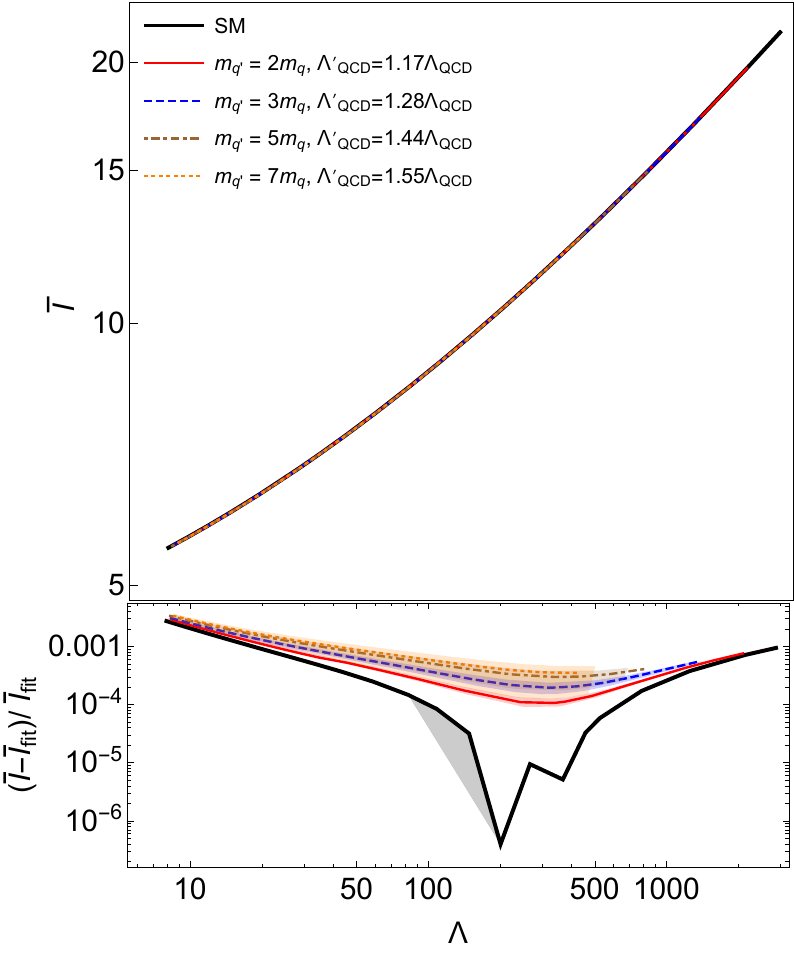}
    \includegraphics[clip=true,width=5.5cm]{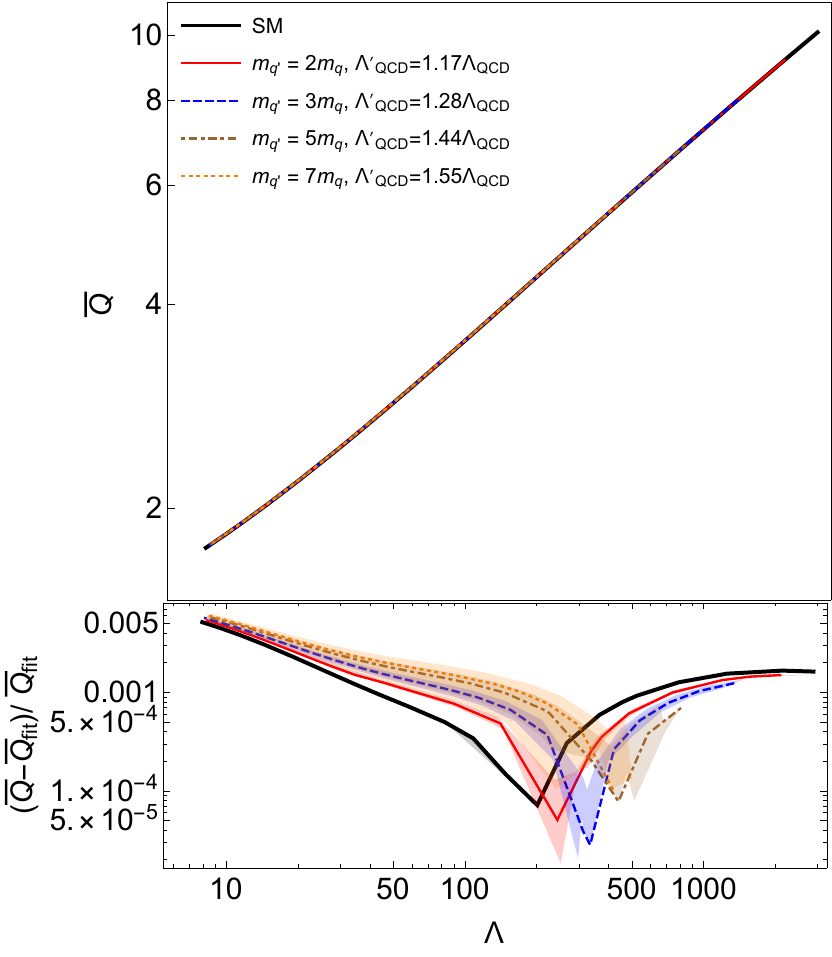}
    \includegraphics[clip=true,width=5.5cm]{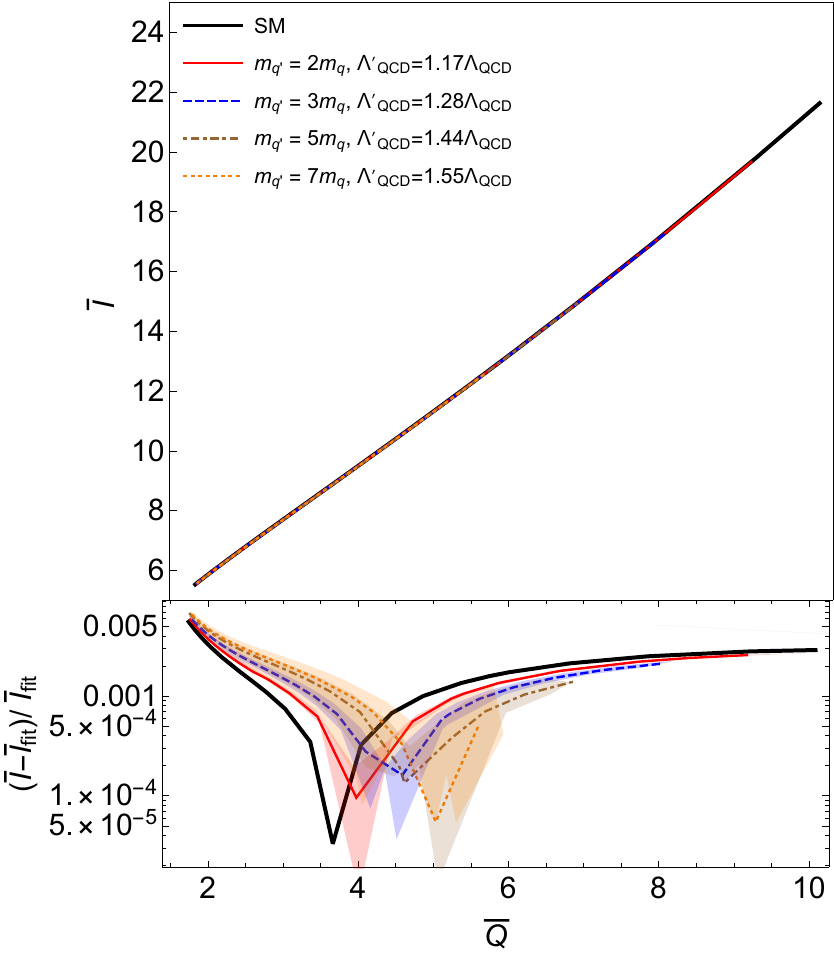}
    \caption{I-Love-Q relations for mirror neutron stars. The top panels show the dimensionless moment of inertia ($\bar{I} = I/M_{*,0}^3$) versus the dimensionless tidal deformability ($\Lambda = \lambda/M_{*,0}^5$) (left), the dimensionless quadrupole moment ($\bar{Q} = Q M_{*,0}/S^2$) versus $\Lambda$ and $\bar{I}$ versus $\bar{Q}$ (right) for a variety of quark mass scalings. The bottom panels show the relative fractional deviation between any relation and a fit to all the data. Observe that the I-Love-Q relations are approximately universal to better than $1\%$, just as in the case of SM neutron stars.
    }
    \label{fig:universal}
\end{figure*}

A natural question is if  certain approximately universal or EoS-insensitive relations~\cite{Yagi:2013bca,Yagi:2013awa,Yagi:2016bkt} continue to hold for mirror neutron stars.  Figure~\ref{fig:universal} show the I-Love-Q relations (top panels) together with deviations from their average (bottom) for different quark mass scalings. It is immediately clear that all of the approximately universal relations satisfied by SM neutron stars are also satisfied by mirror neutron stars. 
The  I-Love-Q relations between moment of inertia, quadrupole moment, and Love number remain ``universal'': they are insensitive to variations in the mirror quark mass to less than 1\%, just like the SM. 

When considering mirror neutron stars in a binary system, we have verified that the inter-relation between the Love numbers in the binary is also insensitive to variations in the mirror quark mass to better than 10\%, also like the SM.  
The same is true for binary Love relations, approximately universal relations between the tidal Love number of mirror neutron stars in a binary system. 

These relations are important because they allow us to break measurement degeneracies. For example, the gravitational waves emitted by a binary system depend on the two tidal deformabilities $\lambda_1$ and $\lambda_2$, which combine into a function $\tilde{\Lambda} = \tilde{\Lambda}(\lambda_1,\lambda_2)$ that can be measured from the data. The binary Love relations allow us to extract $\lambda_1$ and $\lambda_2$ from a measurement of $\tilde{\Lambda}$. Given that the binary Love relations continue to hold in mirror binary neutron stars (to the same level of accuracy as for SM binary neutron stars), the same data analysis pipeline implemented for SM binary neutron stars can also be used to analyze mirror binary neutron stars.

\subsection{Compactness}

\begin{figure*}
    \centering
    \begin{tabular}{cc}
      \includegraphics[clip=true,width=8cm]{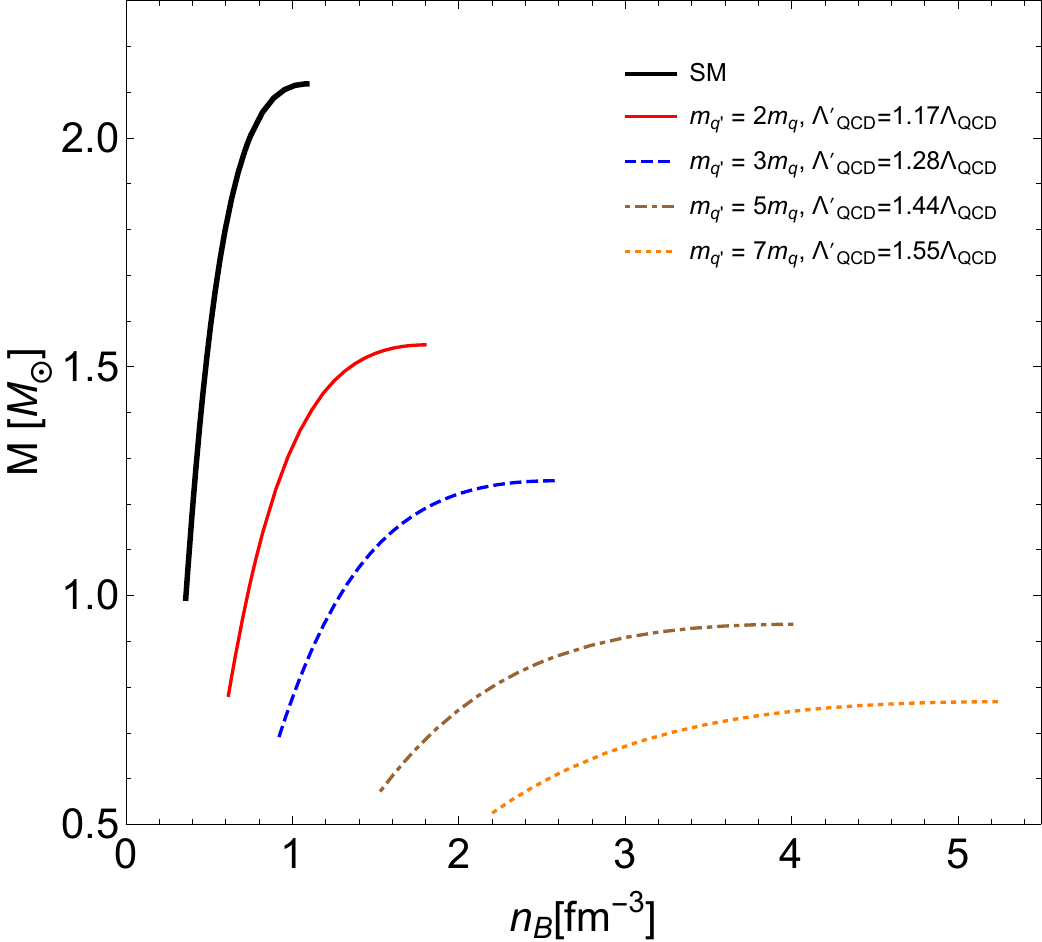}  &   \includegraphics[clip=true,width=8cm]{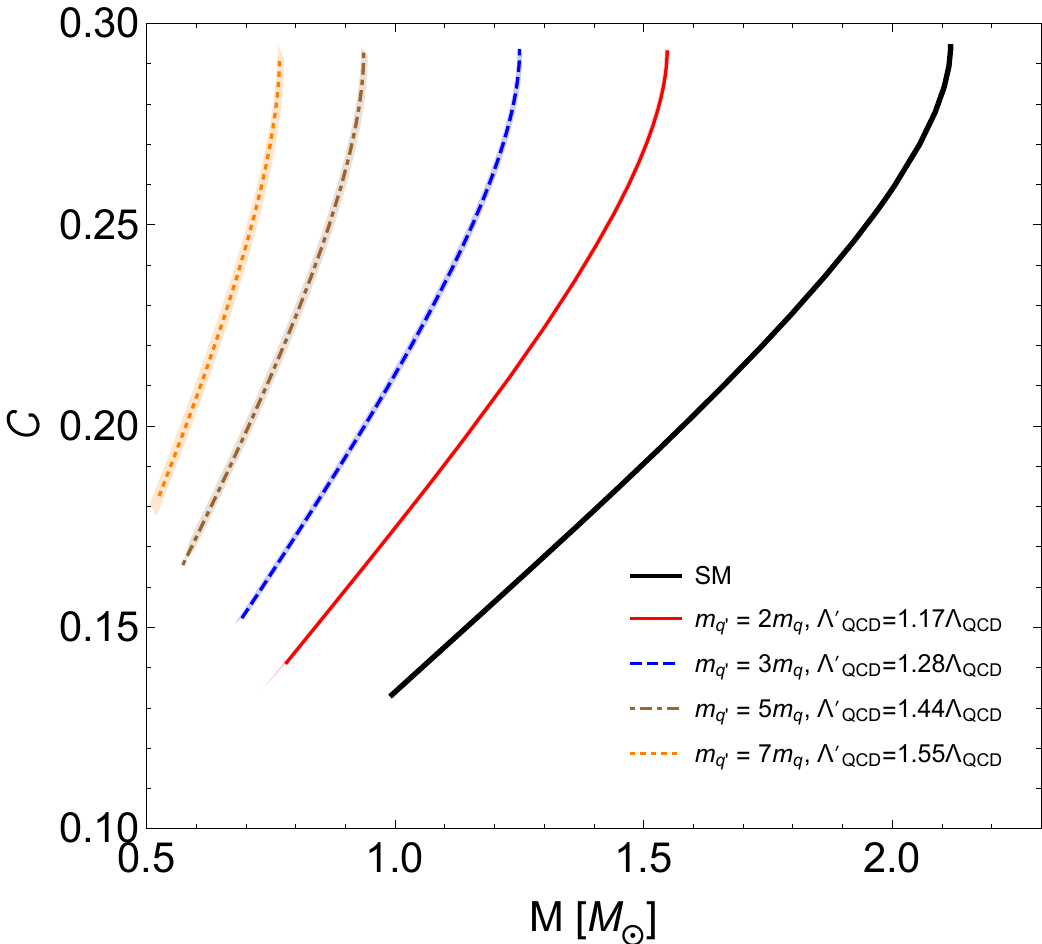}\\
    \end{tabular}
    \caption{Left: Mass of the (mirror matter) neutron star vs. the central baryon density. Right: Compactness of the (mirror matter) neutron star vs. mass.
    }
    \label{fig:nbMplot}
\end{figure*}

As previously discussed, we know that changing quark masses by $f/v$ increases the QCD confinement scale and pion masses, which shifts the EoS to  larger energy densities and higher stiffness, in turn leading to smaller neutron star masses and radii. 
One of the central differences here is that mirror neutron stars are significantly \emph{more} dense at their core than SM neutron stars. The left panel of Fig.\ \ref{fig:nbMplot} shows the mass versus central baryon density for different quark scalings.  The central baryon density for the maximum mass of a mirror neutron star of $m_{q^\prime}=7m_q$ is more than 5 times that of a SM neutron star. While a nuclear physics reader who notes the very large central baryon densities reached in Fig.\ \ref{fig:nbMplot} might wonder about a phase transition into quark matter and/or color superconducting phase, we note that by scaling the quarks masses to higher values this also shift any potential phase transition to higher $n_B$. Thus, we do not anticipate the presence of even a strong first-order phase transition to alter our conclusions.

Since the central density is so much higher than for SM neutron stars, one may wonder whether the compactness ($C = G M/(c^2 R))$ is now close to the black hole limit. For SM neutron stars, $C \in (0.2,0.3)$ approximately, while for a non-rotating black hole $C = 1/2$. The right panel of Fig.\ \ref{fig:nbMplot} shows the compactness for mirror neutron stars as a function of their mass. Interestingly, all mirror neutron stars have the same maximum compactness ($C \lesssim 0.3$), regardless of the $m_{q^\prime}$ scaling. This is completely different from so-called Exotic Compact Objects (ECOs), hypothetical objects that are typically constructed with a surface that is only a small distance from their would be horizons~\cite{Cardoso:2019rvt}. ECOs have compactnesses much closer to the black hole limit than neutron stars, and in this sense, they may act as black hole mimickers. One might have thought that mirror neutron stars would be black hole mimickers, since their central densities can be so large and their radii so small. Yet, we see that they are not, and although they are smaller than neutron stars, their compactness is comparable. 

\section{Observability prospects}
If mirror matter exists, then mirror neutron stars are a natural consequence that leads to an entirely new type of compact object that may be populating our universe, co-existing  with SM neutron stars but completely unconstrained by existing electromagnetic observations. 
Mirror neutron stars can be detected with future gravitational wave observations. The inspiral and coalescence of mirror neutron stars would produce gravitational waves that could be detected by advanced LIGO and its European and Japanese partners. Because of their lower mass, the merger phase would be outside their sensitivity band, but the early and late inspirals would remain in band. The latter could be used to measure the Love numbers through the binary Love relations. A measurement of the masses of the objects, their Love numbers, and lack of electromagnetic signature would then be enough to distinguish between SM neutron stars, primordial black holes, quark stars, and mirror neutron stars.
For example, if advanced LIGO detected the inspiral of a \textit{single} electromagnetically dark binary with masses around $1 M_\odot$ and Love numbers of $\Lambda \sim {\cal{O}}(100)$, this would only be consistent with mirror neutron stars, as it could not be SM neutron star binary or a quark star binary ($\Lambda \sim {\cal{O}}(10^3)$ for this mass~\cite{Yagi:2013bca}), or primordial black holes ($\Lambda = 0$).
Given that future gravitational wave observatories will be sensitive to merger events in the entire observable universe~\cite{Sathyaprakash:2012jk,Reitze:2019iox}, this will provide an extremely powerful probe of dark sector compact objects:  even a non-observation can severely disfavour many new physics scenarios, despite the difficulty of predicting detailed formation rates for mirror neutron stars from first principles. However, it would of course be very interesting to understand the formation rate of mirror neutron star binaries in more detail.

Binary pulsar observations might also detect mirror neutron stars, if asymmetric binaries of a SM pulsar and a mirror neutron star exist and the SM pulsar's radio emissions were observed. 
If the binary is sufficiently relativistic, it may be possible to measure the moment of inertia of its components, because this leads to orbital precession through spin-orbit interactions that impact the timing model. 
Analogously to the measurement of the Love number, the resulting moment of inertia and mass measurements could allow mirror neutron stars to be distinguished from other possibilities for the pulsar's partner.

As discussed above, despite the exotic nature of mirror neutron stars, approximately universal I-Love-Q and Binary-Love relations still hold \cite{Yagi:2013bca,Yagi:2013awa,Yagi:2016bkt} continue to hold for mirror neutron stars.  Figure~\ref{fig:universal}. 
This allows mirror neutron star properties to be extracted from gravitational wave data using standard techniques~\cite{Chatziioannou:2018vzf,Abbott:2018exr}.

A final question would concern the expected detection rate of mirror neutron-star mergers. 
Because the magnitude of the GW signal scales with ${\mathcal{M}^{5/6}}/{D_L}$, where $\mathcal{M}$ is the chirp mass and $D_L$ is the distance to the source, we expect the horizon distance of mirror neutron star mergers to be of about $\sim (\mathcal{M}_\text{mirror}/\mathcal{M}_\text{SM})^{5/6}\sim 0.2 - 0.5$ times that of its SM counterparts. The rate for these events would scale as $\mathcal{R}_\textrm{mirror}\sim \mathcal{R}_\textrm{SM} (R_\textrm{mirror}/R_\textrm{SM})^3\,  \mathcal{F}\sim (0.008 - 0.125)\times \mathcal{R}_\textrm{SM}\times \mathcal{F}$, where  $\mathcal{R}_\textrm{SM}$ is the rate of SM neutron-star mergers, $R_{\textrm{mirror},\textrm{SM}}$ is the horizon distance for mirror and SM neutron-star mergers, while $\mathcal{F}$ is the relative abundance of mirror neutron-star mergers to SM ones. The ratio of horizon distances is simply $R_\textrm{mirror}/R_\textrm{SM}\sim 0.2 - 0.5$, as argued above. The rate $\mathcal{R}_\textrm{SM}$ and the relative abundance $\mathcal{F}$, however, are highly uncertain. They both depend on the specific astrophysical formation channel for these stars, which even in the SM case is uncertain by 2 orders of magnitude~\cite{LIGOScientific:2010nhs,LIGOScientific:2017vwq}.

\section{Robustness of Our Findings}

\subsection{Uncertainties from the Standard-Model Neutron-Star Equation of State}

Because the mass-radius relations for SM neutron stars is not known with certainty, one might ask how uncertainties in the SM neutron-star EoS affect our results.   
The effect of these uncertainties can be estimated by employing the scaling of the mirror star masses and radii with the mirror baryon mass $m_B'$. We note that, because the mass and radius of mirror neutron stars scale with a negative power of the baryon mass, $\propto (m_B'/m_B)^{-1.9}$, uncertainties from the SM mass-radius relation tend to become \emph{smaller} when extrapolated to the mirror sector. By rescaling the uncertainty in the radius of a Chandrasekhar-mass neutron-star in this manner, we have checked that the expected mass-radius ranges for standard-model and mirror neutron stars \emph{do not} overlap in the phenomenologically relevant range of $m_q'/m_q \gtrsim 2$. Because gravitational-wave signatures are tightly related to the mass and radius of the merging objects, this indicates that the distinctive features of mirror neutron-star mergers are robust against uncertainties from the standard-model EoS.   Moreover, as new observations lead to tighter constraints for the mass-radius relation of ordinary neutron stars, mirror neutron-star mergers should become increasingly distinguishable from their standard-model counterparts. 

Finally, our core model does not incorporate the possibility of hyperon nor quarks degrees of freedom.  While this would change the general shape of the SM mass-radius relationship \cite{Alford:1997zt,Alford:1998mk,Alford:2001dt,Buballa:2003qv,Alford:2004pf,Fukushima:2003fw,Alford:2006vz,Alford:2002rj,Alford:2007xm,Dexheimer:2009hi,Dexheimer:2011pz,Alford:2013aca,Baym:2019iky,Annala:2019puf,Tan:2020ics} we do not anticipate any significant effect in our scalings because these degrees of freedom would also scale with $f/v$.

\subsection{Sensitivity to Mirror Matter Model}

One might also question the extent to which our results are sensitive to the particular mirror-sector model employed in our calculations---namely, the $Z_2$-symmetric Mirror Twin Higgs, where all mirror quark masses equal the standard-model ones times a factor of $f/v$. It is possible that the Mirror Twin Higgs scenario is realized in nature in a different manner. For example, hard $Z_2$-breaking Yukawa couplings for the light mirror quarks are frequently considered in the literature (see e.g.~\cite{Barbieri:2016zxn}), and would lead to different masses for the mirror quarks, baryons and mesons. Nevertheless, as long as these changes do not qualitatively change the character of the mirror hadron sector, we expect the $\sim m_{B}'^{-2}$ scaling of the mass-radius relation to persist as a reasonable approximation. In that case, the mass-radius relation of mirror neutron stars can be readily estimated from the dark baryon mass. However, this might not be the case if the dark QCD sector is qualitatively different. For instance, if the dark quarks are much heavier than the $Z_2$-symmetric expectation, such that the number of mirror quark flavors $N_f$  with mass below the dark confinement scale is smaller than 2, then the character of dark QCD and the mirror baryon dark matter component would change qualitatively due to the absence of dark pions for $N_f = 1$, or the qualitatively different origin of baryon masses for $N_f = 0$. In those cases, a new analysis of mirror neutron stars would be necessary.

\subsection{Uncertainties in Mirror-Sector Scalings}
\label{sec:scalinguncertainty}


The shaded bands in Fig.~\ref{fig:EoS} correspond to the uncertainty in the pion mass dependence of the EoS input parameters, obtained from lattice data and chiral perturbation theory. The relative error grows with increasing quark mass, reflecting the fact that the further our mirror model is deformed from SM QCD, the greater the uncertainty in the properties of mirror QCD resonances. Note that the $p-\epsilon$ curve shifts to larger energy densities and pressures for increasing quark mass. This is because, to a good approximation, the pressure and energy density increase in proportion to $(\Lambda_{QCD}')^4$. 
This scaling is in agreement with the naive expectation, since both the mass of the neutron and nuclear saturation density are both (to zeroth order in $m_\pi^2 / \Lambda_{QCD}'^2$) increasing with the appropriate powers of $\Lambda_{QCD}'$.

We have further demonstrated that our results are insensitive to the fine details of the scaling of the nuclear binding energies with quark mass. Multiplying all nuclear binding energies by random numbers in the range $(0.5, 2)$ has only a percent-level effect on our results, showing that our crust model is robust with respect to these unknown parameters.

There are three inputs to the core model which we do not rescale with the quark mass: $a_3$, $a_4$, and $g_{\omega\rho}$, which premultiply quartic interactions of the sigma, omega and rho mesons. We do not expect rescaling of these parameters to have a significant effect on our conclusions (see above); nonetheless lattice data for the pion mass dependence of these couplings would provide information that our core model currently lacks, and would be especially valuable insofar as they would render our predictions more robust. Similarly, the other parameter that does not get rescaled with quark mass is $r_b$, which determines the density at which our interpolation function is matched to the core model. Improving on this would require a more detailed model of the regime between the core and crust, and a model therefore of mirror nuclear pasta and liquid nuclear matter, which is beyond the scope of this work. However, we do not anticipate significant effects with their inclusion beyond subtle differences in the low mass radius \cite{Gamba:2019kwu}.


\subsection{Mass-Radius Scaling Relations}

A strong argument for the robustness of our conclusions lies in how the mass-radius relation scales with the mirror baryon mass: $M$, $R\sim {m_B'}^{-1.9}$, with interactions slightly shifting the exponent from the power of $-2$ expected from dimensional analysis \cite{Narain:2006kx}. Because $m_B'/m_B<1$, this scaling has the effect of reducing uncertainties from the standard-model neutron-star mass-radius relation, when these are propagated to the mirror sector. It also controls systematic uncertainties from the scaling of model parameters: as long as $m_B'$ is scaled correctly, results should fall in the right ballpark.  

\begin{figure}
    \centering
    \includegraphics[width=8cm]{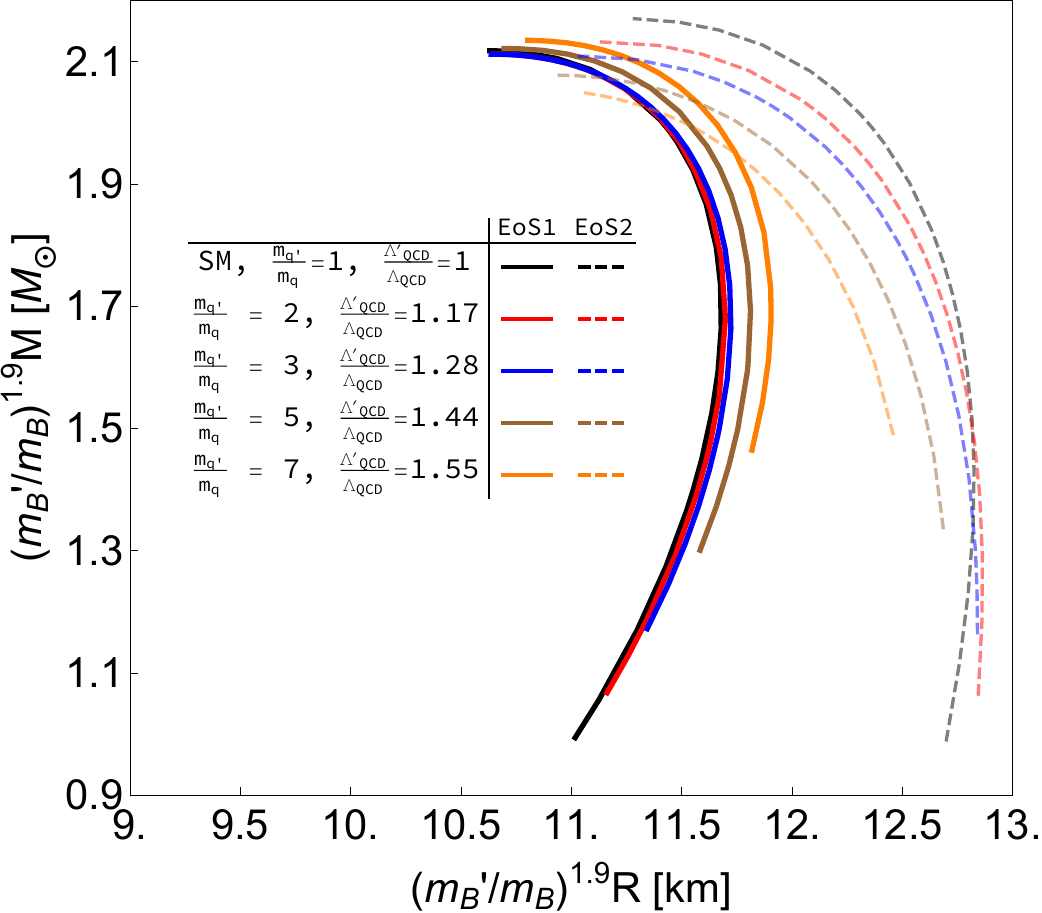}
    \caption{
    Scaling of the mass-radius relation with $(m_B'/m_B)^{-1.9}$.
    Solid lines (EoS1) represent results for the EoS from Sec.~\ref{sec:EoS}. 
     The shaded regions around these curves represent uncertainties in the $m_\pi$ dependence of various EoS input parameters (from chiral perturbation theory and Lattice QCD). Faint lines (EoS2) correspond to the modified EoS from Sec.~\ref{sec:altEoS}. The distance between faint (EoS2) and solid (EoS1) curves is meant to roughly represent systematic uncertainties.
    }
    \label{fig:scaling}
\end{figure}

In broad terms we should expect the two important dimensionful parameters for the physics of neutron stars to be $m_n$ and the Planck mass. Since the Planck mass is constant, we would naively expect the mass and radius of a neutron star to scale with the neutron mass as $1/m_n^2$~\cite{Reisenegger:2015crq}.  
In fact we find that our complete numerical solutions for neutron star structure closely reflect this naively expected scaling relation. We can reproduce our numerical results to a reasonable approximation by simply rescaling the mass and radius of SM neutron stars by $(m_n'/m_n)^d$, with $d \approx -1.9$, see Fig.~\ref{fig:scaling}. Furthermore the scaling exponent $d$ does not appear to change for different choices of the parameters $a_3$, $a_4$, $g_{\omega\rho}$, strongly indicating that this general behaviour is robust, even beyond the finer details of our model, as long as these parameters do not themselves have a strong dependence on the pion mass. An equally good scaling of the mass-radius relations can be found by considering mirror neutron stars to be interaction dominated and rescaling $M$ and $R$ by $(m_n\, m_\omega/g_\omega)^{-1}$ or by $(m_n\,f_\pi)^{-1}$ \cite{Narain:2006kx}. 

Based on the robust dependence of our results on the mirror confinement scale / neutron mass, we also predict that hidden sectors with lower $\Lambda_\mathrm{QCD}'$ than the SM (e.g. due to mirror quarks lighter than the SM) would give rise to mirror neutron stars with higher masses, Love numbers, and moments of inertia than SM neutron stars. Although this scenario is not motivated by the Mirror Twin Higgs model that we study, it could arise in more general dark sector models and models of Dark Complexity.

Scaling relations for the masses and radii, for different values of the (mirror) quark mass $m_q'$, are shown in Fig.~\ref{fig:scaling}.  In fact, one obtains roughly the same mass-radius relations for all values of $m_q'/m_q$  under all scenarios, as long as one rescales both masses and radii by $(m_B'/m_B)^{1.9}$. While the scaling with $m_B'$ to power $-1.9$ becomes slightly worse with modifications to the equation of state (see Sec.~\ref{sec:altEoS}), it still remains valid within less than 10\%.  Hence, we find that, while the particular shape of the mass-radius relation can be sensitive to model details and to imprecision in the extraction of parameter scalings from lattice QCD, these details are not sufficiently important to make a difference in the overall scales for mirror neutron stars, and do not alter our main conclusions.

\subsection{Robustness Check: Modified Equation of State}
\label{sec:altEoS}

To explicitly check that the uncertainties discussed above do not have a strong influence on our results, we calculate the effect of modifying the quark-mass scalings, particle content, and assumptions in our model. 
We note that with these modifications we are purposely trying to take \emph{very drastic changes} to the EOS in order to see if we can break the mass-radius scalings.  
For instance, we completely remove protons and electrons such that the stars only contain neutrons, which would be a very extreme limit. 
Thus, these modifications provide a very strong test of the robustness of our results.

In practice, we:
    \begin{enumerate}
        \item Scale the value of saturation density in the mirror sector with the baryon mass $m_B'$, instead of $\Lambda_{\textrm{QCD}}'$, as $n_{\textrm{sat}}' \propto {m_B'}^3$.
        \item Remove the scaling of the coupling to the $\sigma$ and $\omega$ mesons, $g_\sigma$ and $g_\omega$, leaving them unchanged in the mirror sector (same as standard-model value).
        \item Use the scaling originally intended for the coupling to the $\sigma$ meson, $g_\sigma$, for the $\sigma^4$-coupling, $a_4$, which was originally left unchanged in the mirror sector.
        \item Use the scaling originally intended for the coupling to the $\rho$ meson, $g_\rho$, for the $\sigma^3$-coupling, $a_3$, which was originally left unchanged in the mirror sector.
        \item Decouple the $\rho$ meson from our model altogether by setting its standard-model coupling $g_\rho=0$.
        \item Consider pure neutron matter, setting the fraction of protons to zero.
    \end{enumerate}


The results of these modifications are represented by the faint lines labeled EoS2 in Figs.~\ref{fig:MRplotfront}, \ref{fig:mirrorobservablesplot} and \ref{fig:scaling}, where each color represents a different value of $f/v=m_q^\prime/m_q$. In particular, Fig.~\ref{fig:scaling} shows that the scaling of the mass-radius curves with the quark mass still holds within $\sim 10\%$ for the modified set of EoSs. 
Moreover, because stellar masses and radii decrease with increasing mirror baryon mass $m_B'$, one finds that the absolute difference between strong and faint lines in  Figs.~\ref{fig:MRplotfront} and \ref{fig:mirrorobservablesplot} actually becomes \emph{smaller} for mirror neutron stars, especially at larger values of $f/v$. That is, because of the scaling with $m_B'$ uncertainties do not worsen, but rather improve for heavier mirror particles, at least in absolute terms.

 While the modified EoS leads to a new shape of the mass-radius relation, approximately straight up and down with a large radius at low masses, stellar masses and radii are still in the same ballpark, and remain distant from the region for standard-model neutron stars. This means our conclusions are robust, even under extreme modifications to our model. 


\section{Conclusions}

We have shown that mirror neutron stars predicted by the Mirror Twin Higgs solution to the hierarchy problem can be detected in standard LIGO observations and analyses.  The detection of mirror neutron stars would revolutionize cosmology, particle physics and nuclear physics. It would answer a fundamental question concerning the nature of dark matter. It would prove the existence of hidden sectors and allow us to investigate their possible connection to other mysteries like the hierarchy problem and baryogenesis. Finally, it would lead to a new frontier in nuclear physics, providing an alternative laboratory for probing nuclear matter with different fundamental properties. This is particularly exciting because Lattice QCD calculations can be performed across a range of quarks masses, and are in fact easier to perform for heavier (mirror) pions. Such a discovery would enrich our understanding of confining gauge forces and their bound states to an inestimable degree. 
Mirror neutron stars represent an exciting opportunity for gravitational wave and binary pulsar observations to make ground-breaking fundamental discoveries at the heart of several interlinked physics disciplines.

{\bf \emph{Acknowledgements} -- } 
The authors would like to thank Veronica Dexheimer for useful discussions related to this work. 
M.H. thanks J\"urgen Schaffner-Bielich for insightful comments. 
H.T. and N.Y.~acknowledge financial support through NSF grants No.~PHY-1759615, PHY-1949838 and NASA ATP Grant No.~17-ATP17-0225, No.~NNX16AB98G and No.~80NSSC17M0041. 
The research of J.S. and D.C. was supported in part by a Discovery Grant from the
Natural Sciences and Engineering Research Council of Canada, and by the Canada Research Chair
program. J.S. also acknowledges support from the University of Toronto Faculty of Arts and Science postdoctoral fellowship. 
J.N.H. acknowledges financial support by the US-DOE Nuclear Science Grant No. DESC0020633.  
This work was supported in part by the National Science Foundation (NSF) within the framework of the MUSES collaboration, under grant number OAC-2103680.
\bibliography{references2, noninspire}


\end{document}